\documentclass[a4paper,fleqn]{cas-sc}
\usepackage{bm}
\usepackage[numbers]{natbib}
\usepackage{amsmath}
\usepackage{graphicx}  
\usepackage{float}  
\usepackage{subfigure}

\usepackage{algorithm}
\usepackage{algorithmic}
\usepackage{appendix}

\usepackage{amsmath,amssymb,amsfonts}
\usepackage{textcomp}
\usepackage{wrapfig}
\usepackage{stfloats}

\usepackage{bbm} 
\usepackage{color}
\usepackage{setspace}
\graphicspath{
	{fig3/}
}

\def\tsc#1{\csdef{#1}{\textsc{\lowercase{#1}}\xspace}}
\tsc{WGM}
\tsc{QE}

\begin{document}
	\let\WriteBookmarks\relax
	\def\floatpagepagefraction{1}
	\def\textpagefraction{.001}
	
	
	\title [mode = title]{Dynamic Offloading Loading Optimization in distributed Fault Diagnosis system with Deep Reinforcement Learning Approach}  


%

\author[1]{Liang Yu}
\author[2]{Qixin Guo}
\author[2]{Rui Wang}
\cormark[1]
\ead{ruiwang@tongji.edu.cn}
\author[2]{Minyan Shi}
\author[3]{Fucheng Yan}
\author[3]{Ran Wang}






\affiliation[1]{organization={Instutute of Vibration, Shock and Noise, State Key Laboratory of Mechanical System and Vibration},
	addressline={Shanghai Jiao Tong University},
	city={Shanghai},
	postcode={200240},
	country={China}}
\affiliation[2]{organization={College of Electronics and Information Engineering},
	addressline={Tongji University},
	city={Shanghai},
	postcode={201804},
	country={China}}
\affiliation[3]{organization={College of Logistic Engineering},
	addressline={Shanghai Maritime University},
	city={Shanghai},
	postcode={201306},
	country={China}}
\cortext[cor1]{Corresponding author}

\begin{abstract}
Artificial intelligence and distributed algorithms have been widely used in mechanical fault diagnosis with the explosive growth of diagnostic data.
A novel intelligent fault diagnosis system framework that allows intelligent terminals to offload computational tasks to Mobile edge computing (MEC) servers is provided in this paper, which can effectively address the problems of task processing delays and enhanced computational complexity.
As the resources at the MEC and intelligent terminals are limited, performing reasonable resource allocation optimization can improve the performance, especially for a multi-terminals offloading system. 
In this study, to minimize the task computation delay, we jointly optimize the local content splitting ratio, the transmission/computation power allocation, and the MEC server selection under a dynamic environment with stochastic task arrivals. 
The challenging dynamic joint optimization problem is formulated as a reinforcement learning (RL) problem, which is designed as the computational offloading policies to minimize the long-term average delay cost. 
Two deep RL strategies, deep Q-learning network (DQN) and deep deterministic policy gradient (DDPG), are adopted to learn the computational offloading policies adaptively and efficiently. 
The proposed DQN strategy takes the MEC selection as a unique action while using the convex optimization approach to obtain the local content splitting ratio and the transmission/computation power allocation. 
Simultaneously, the actions of the DDPG strategy are selected as all dynamic variables, including the local content splitting ratio, the transmission/computation power allocation, and the MEC server selection. 
Numerical results demonstrate that both proposed strategies perform better than the traditional non-learning schemes. 
The DDPG strategy is superior to the DQN strategy due to its ability to learn all variables online, although the DDPG strategy requires large complexity.
\end{abstract}


\maketitle
\section{Introduction} 
Large-scale and integrated equipment puts forward higher requirements for condition monitoring with the improvement of productivity \cite{zhang2019privacy, yang2022multi, azamfar2020intelligent}. 
Intelligent mechanical fault diagnosis algorithms have been accompanied by the development of artificial intelligence (AI) and Internet of Things (IoT) technologies, such as the application of deep learning (DL) and reinforcement learning (RL) in fault diagnosis \cite{aslam2020internet, yoo2022vibration, lei2020applications, li2022perspective, wang2022semi, chen2022residual}.
A collaborative deep learning-based fault diagnosis framework is proposed to solve the data transmission problem in distributed complex systems, which is a security strategy that does not require the transmission of raw data \cite{wang2021collaborative}.
An Improved classification and regression tree algorithm are proposed, which ensures the accuracy of fault classification by reducing the iteration time in the computation \cite{deng2020high}.
A fault diagnosis method based on adaptive privacy-preserving federated learning is used for the Internet of Ships, which guarantees no risk of data leakage by sharing model parameters \cite{zhang2021adaptive}.
A deep learning-based approach to automated fault detection and isolation is used for fault detection in automotive dashboard systems, which is tested against data generated from a local computer-based manufacturing system \cite{iqbal2019fault}.
An intelligent fault detection method based on the multi-scale inner product is adopted for shipboard antenna fault detection, which uses the inner product to capture fault information in vibration signals and combines it with locally connected feature extraction \cite{pan2019novel}.


\begin{figure}[!htb]
	\centering
	\includegraphics[width=0.8\textwidth]{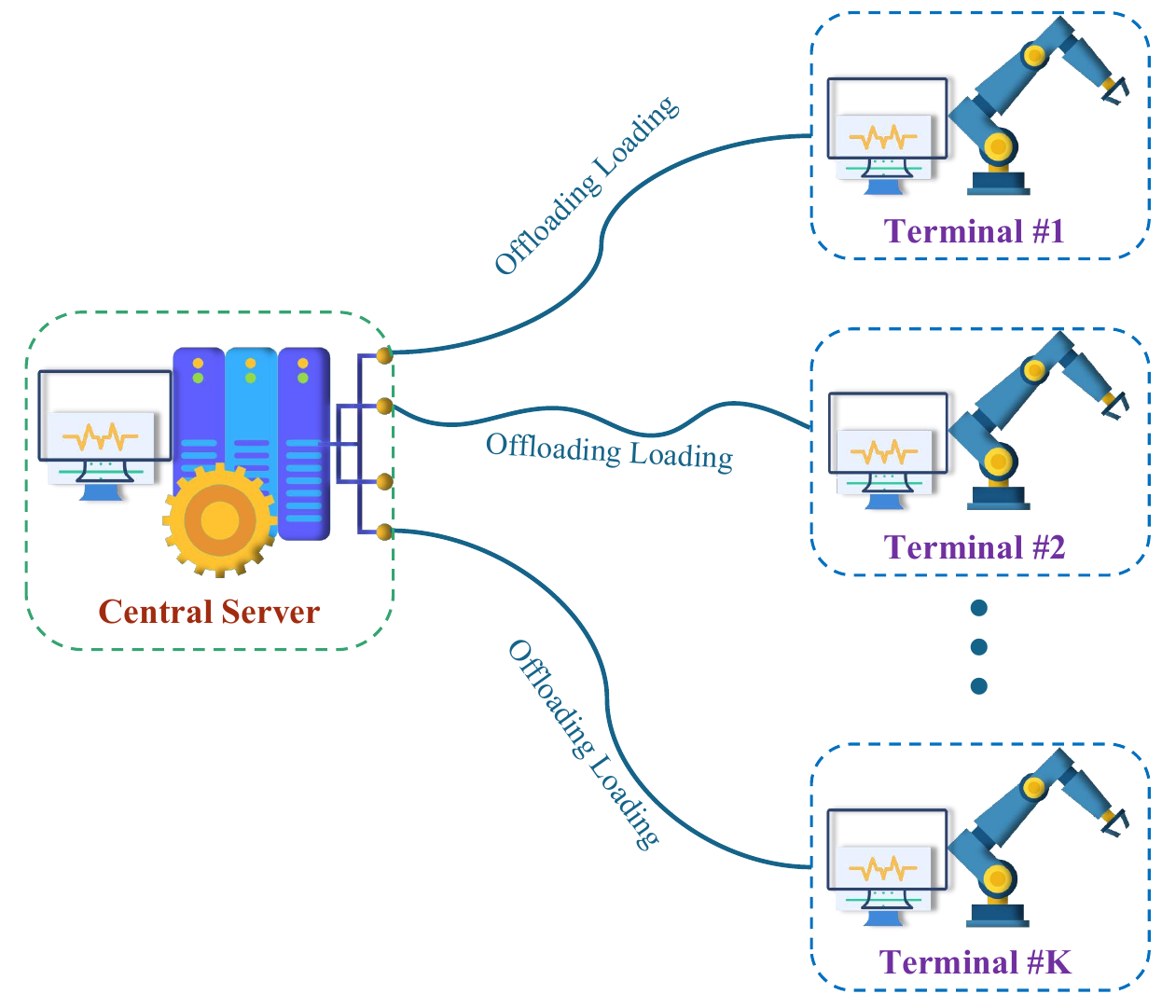}
	\caption{The framework of the conventional mechanical fault diagnosis system, in which the terminal uploads the monitoring data to a central server through the network cable for processing. The central server has powerful computing power but is generally far away from the terminal. The terminal is only responsible for collecting monitoring data and typically has no computing power.}
	\label{Network_model1}
\end{figure}


The current intelligent fault diagnosis algorithm pays more attention to the reliability of the diagnosis and less attention to the timeliness \cite{liu2020reliability}. 
The server's computation resources and the timeliness of data processing have become urgent problems to be solved with the exponential growth of diagnostic data throughput.
The traditional fault diagnosis systems offload the diagnostic data collected by terminals to a server with powerful computing power for processing, as shown in Fig. \ref{Network_model1}.
The server is usually far away from the acquisition terminal, which causes a waste of resources during transmission and increases data transmission delay \cite{kumar2013survey}.
The emergence of mobile edge computing (MEC) provides a solution to these problems, which is considered a promising architecture for data access \cite{mao2017survey, huda2022survey, liao2022online}.
MEC deploys several lightweight servers closer to the collection terminals compared to traditional state monitoring systems, which are called mobile edge servers.
MEC servers can reduce the burden of performing computation for large content tasks and task processing delays significantly by allowing terminals to offload computation tasks to a nearby MEC server \cite{lu2022secure, guo2022distributed}.

\begin{figure}[!htb]
	\centering
	\includegraphics[width=0.85\textwidth]{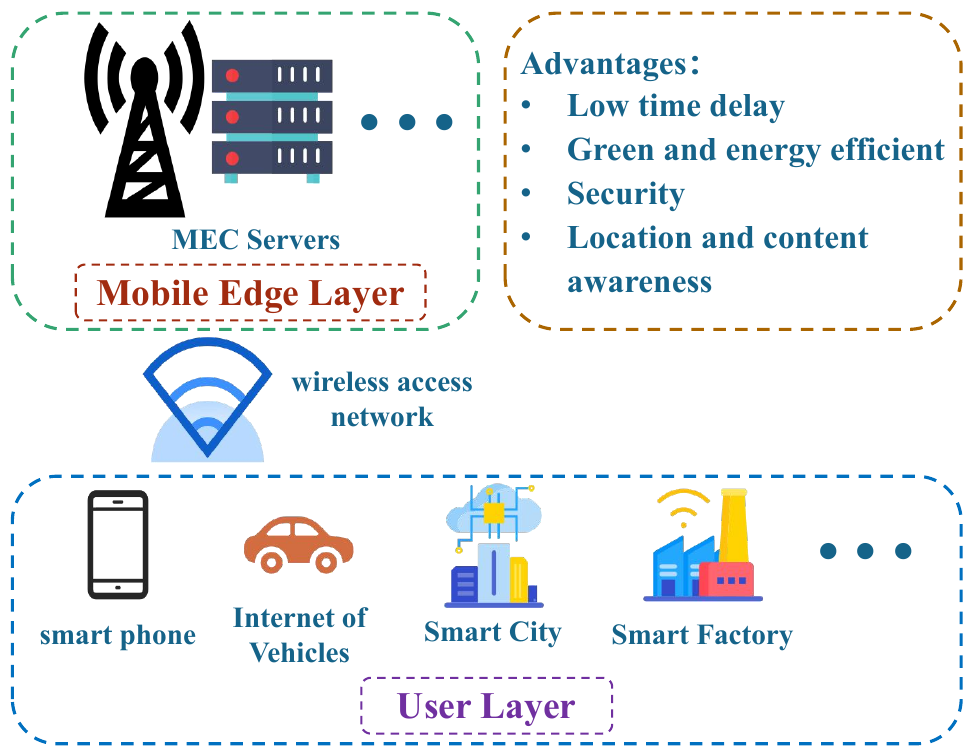}
	\caption{Architecture, applications, and advantages of MEC. The user layer is composed of mobile device terminals with certain computing capabilities, which have a wide range of application scenarios in the IoT, such as 5G communication, virtual reality, Internet of Vehicles, smart city, smart factory, etc. The mobile edge layer consists of edge servers close to the device terminals with high computing capabilities. The interaction between the user layer and the mobile edge layer is completed in the framework of the wireless access network.}
	\label{fig:2}
\end{figure}

The architecture of MEC usually consists of the user layer and the mobile edge layer \cite{esposito2017challenges, liu2020toward, wu2019ledge, cui2020online}, as shown in Fig. \ref{fig:2}.
In the MEC paradigm, the user layer consists of mobile device terminals, which contain various applications and functions and also have certain computing capabilities.
When processing each computing task, the device terminal can choose to process it on its own device in addition to offloading the task to the mobile edge layer or cloud layer through data transfer.
The mobile edge layer consists of edge servers near the device terminals, which computing resources are more abundant than those of the device terminals.
Through computing offload technology, information can be interacted with in real-time to meet the computing needs of different types of application scenarios. 
The MEC architecture has a wide range of application scenarios in the IoT, such as 5G communication, virtual reality, Internet of Vehicles, smart city, smart factory, etc.
The MEC architecture has the advantages of low time delay, green and energy efficiency, security, location, content awareness, etc., which makes it easier to access AI methods and blockchain methods.

Computing offloading as one of the core techniques of the MEC has received great attention recently. 
For simple, indivisible, or highly integrated tasks, binary offloading strategies are generally adopted, and tasks can only be computed locally or all offloaded to the servers \cite{mao2016dynamic}. 
The authors in \cite{barbarossa2014communicating} formulated the binary computation offloading decision problem as a convex problem, which minimizes the transmission energy consumption under the time delay constraint. 
The computation offloading model studied in \cite{zhang2013energy} assumed that the application has to complete the computing task with a given probability within a specified time interval, for which the optimization goal is the sum of local and offloading energy consumption. 
This work concluded that offloading computing tasks to the MEC servers can be more efficient in some cases.
In practice, offloading decisions can be more flexible. The computation tasks can be divided into two parts performed in parallel: one part is processed locally, and the other is offloaded to the MEC servers for processing \cite{zhang2012offload}. 
A task-call graph model is proposed to illustrate the dependency between the terminal and MEC servers, in which decisions and latencies are investigated by the joint offloading scheduling and formulated as a linear programming problem \cite{mahmoodi2016optimal}.

\begin{figure}[!htb]
	\centering
	\includegraphics[width=0.99\textwidth]{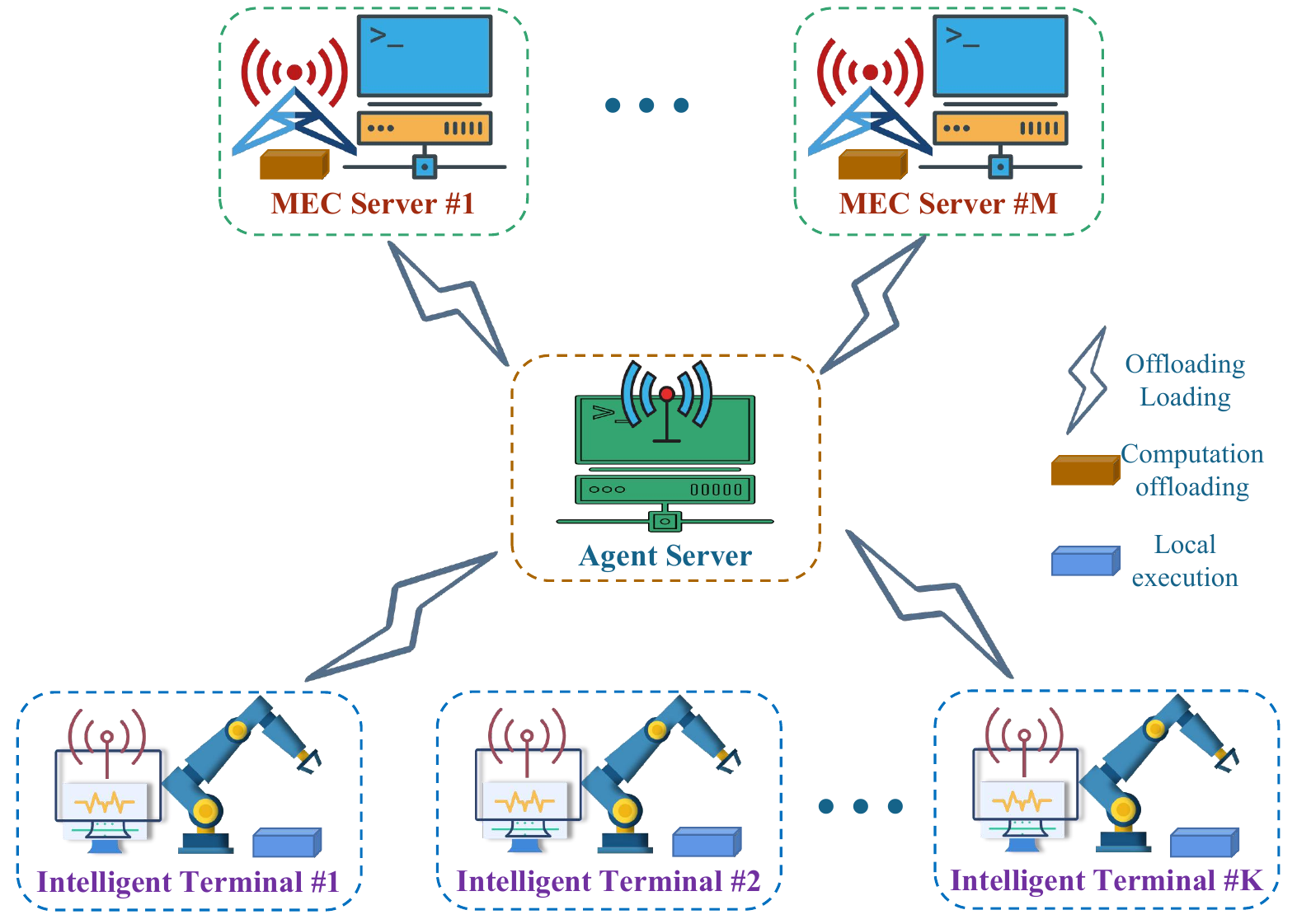}
	\caption{The framework of the intelligent mechanical fault diagnosis system in this paper, whose contains three parts: intelligent terminal, agent server, and MEC servers. The intelligent terminal is responsible for collecting fault diagnosis data and has a weak data processing capability. The MEC servers are small servers with certain data processing capabilities. The agent server is responsible for policy formulation and controls the ratio of intelligent terminals and MEC servers to process the fault diagnosis data.}
	\label{Network_model}
\end{figure}

RL has been employed as a new solution to the problem of MEC offloading, which is a model-free machine learning algorithm that can perform self-iterative training based on the data it generates  \cite{lu2020optimization, wang2020reinforcement, zhao2020deep, ren2020deep}.
Task processing delay is a vital optimization parameter for time-sensitive systems. 
The authors studied the problem of computation offloading in an IoT network in \cite{min2019learning}, in which the Q-learning-based RL approach was proposed for an IoT device to select a proper device and determine the proportion of the computation task to offload.
The authors in \cite{hu2018mobility} investigated joint communication, caching, and computing for vehicular mobility networks. A deep Q-learning-based RL with a multi-timescale framework was developed to solve the joint online optimization problem.
In \cite{wei2018dynamic}, the authors studied the offloading for the energy harvesting (EH) MEC network. An after-state RL algorithm was proposed to address the large time complexity problem, and polynomial value function approximation was introduced to accelerate the learning process.
In \cite{zhang2020dynamic}, the authors also studied the MEC network with the EH device. The authors proposed hybrid-based actor-critic learning for optimizing the offloading ratio, local computation capacity, and server selection. From the above references, efficient computational offloading decisions based on RL methods can help the system to reduce computational complexity and computational time cost.

In the framework of the intelligent fault diagnosis system proposed in this paper, the user layer consists of intelligent terminals with certain computing power, and the mobile edge layer consists of MEC servers with strong computing power, as shown in Fig. \ref{Network_model}. 
The intelligent terminal offloads the fault diagnosis data to any MEC server proportionally through the agent server's policy.
The optimization problem becomes an offloading decision problem in a dynamic MEC environment, and the current channel state information (CSI) cannot be observed while making the offloading decision. 
The offloading policy should follow the predicted CSI and task arrival rates under the intelligent terminal and MEC server energy constraints, aiming to minimize the long-term average delay cost. 
We first establish a low-complexity deep Q-learning network (DQN) based offloading framework where the action includes only discrete MEC server selection, while the local content splitting ratio and the transmission/computation power allocation are optimized by the convex optimization method. 
Then we develop a deep deterministic policy gradient (DDPG) based framework which includes both discrete MEC server selection variable and constant local content splitting ratio, the transmission/computation power allocation variable as actions.
The numerical results demonstrate that both proposed strategies perform better than the traditional non-learning scheme. 
The DDPG strategy is superior to the DQN strategy as it can online learn all variables.
Compared with the traditional fault diagnosis system, the intelligent fault diagnosis system migrates the original computing tasks based on the central server to the edge computing system, which reduces the computing load of the central server, slows down the network bandwidth pressure, and improves the real-time data interaction.
On the other hand, the new intelligent fault diagnosis system solves the problem of the single function of traditional instrumentation systems, which increases the intelligence of instrumentation and makes it easier to access other intelligent methods.

The contributions of this paper can be summarized as follows.

1) A new framework for the intelligent fault diagnosis system based on the MEC framework is proposed, in which MEC servers and intelligent terminals can process monitoring data and the ratio determined by the offload policy of the agent server.
Compared with the traditional fault diagnosis system, the intelligent fault diagnosis system solves the problems of limited computing resources and network delay and increases the intelligence of the equipment. 

2) Two offloading scenarios of the intelligent fault diagnosis system are modeled: one-to-one and one-to-multiple. 
One-to-one means that one MEC server can only be connected by one intelligent terminal simultaneously, and one-to-multiple implies that multiple intelligent terminals can be connected to the same MEC server simultaneously. 
The optimization goal is taking the maximum time delay for the system to complete the computation task at each time slot.
Every intelligent terminal and MEC server has its energy constraints, and the agent determines the power allocation during the offloading process.

3) The offloading decision optimization algorithm based on the combination of convex optimization and deep reinforcement learning is designed. 
Firstly the convex optimization methods are used to solve the connection problem of the intelligent terminal need to choose which MEC server. 
Then the resource allocation of intelligent fault diagnosis system offloading is given by DQN and DDPG algorithm.

The remainder of this paper is structured as follows. The intelligent fault diagnosis system models are provided in Section 2. The DDPG-based Offloading Design and DQN-based Offloading Design are described in Section 3 and Section 4, respectively. The Numerical results and relevant analysis are presented in Section 5. The conclusion is given in Section 6.


\section{The Intelligent Fault Diagnosis System Model}
\begin{figure}[!htb]
	\centering
	\includegraphics[width=0.99\textwidth]{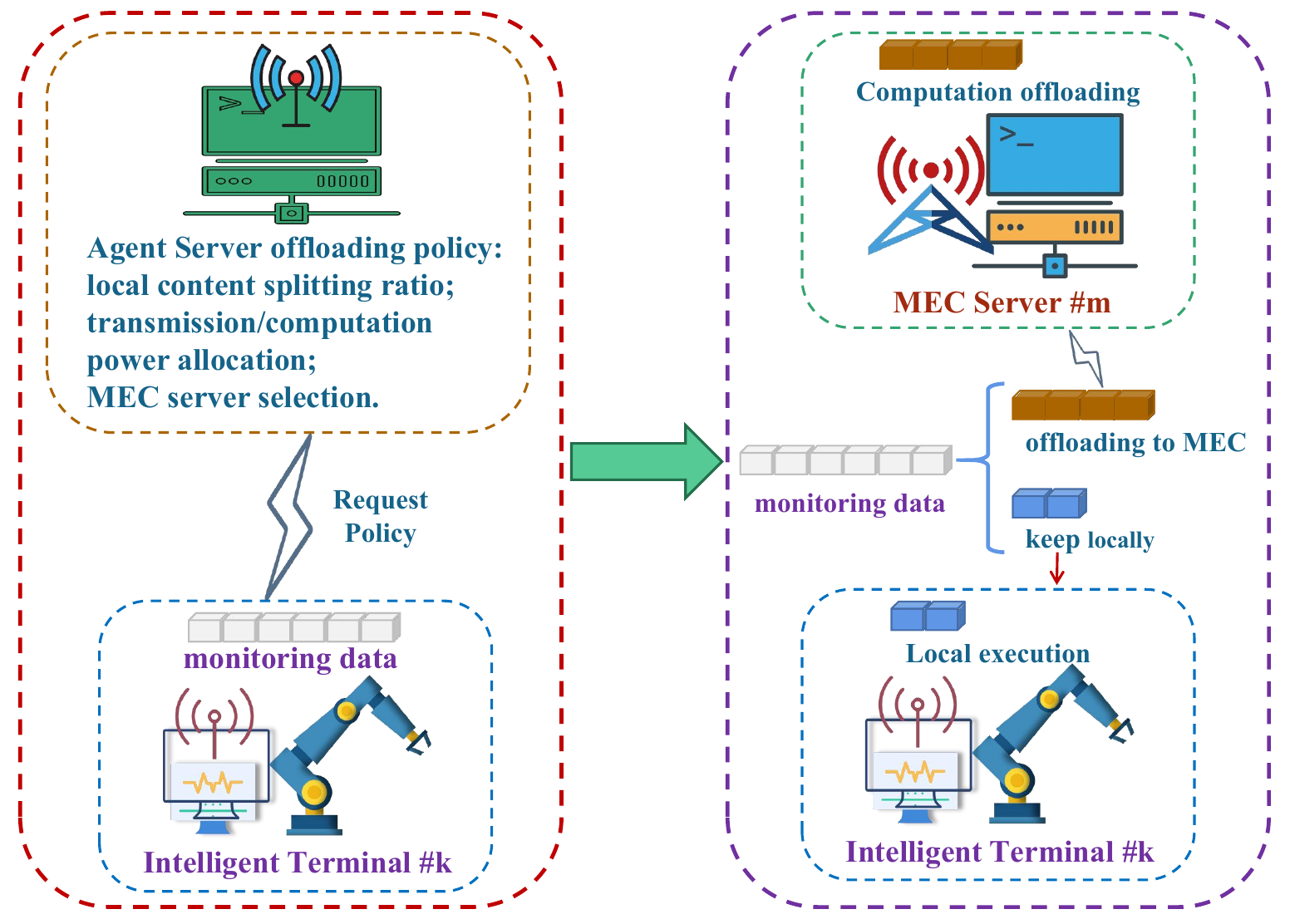}
	\caption{The working principle of intelligent fault diagnosis system in this paper. The intelligent terminal collects the fault diagnosis data and then requests a policy from the agent server. The offload policy of the agent server includes the local content splitting ratio, the transmission/computation power allocation, and the MEC server selection. Finally, the agent server offloads the fault diagnosis data to the MEC server according to the ratio determined by the offload policy.}
	\label{fig:1}
\end{figure}

\
A new framework for the intelligent fault diagnosis system is proposed in this paper, which consists of MEC servers and intelligent terminals, as shown in Fig. \ref{fig:1}.
Both MEC servers and intelligent terminals can process monitoring data, and the intelligent terminal can offload data to any MEC server through the agent. 
The interaction between the intelligent terminal and the MEC server operates in the orthogonal frequency division multiple access frameworks.
The offloading policy includes the local content splitting ratio, the transmission/computation power allocation, and the MEC server selection.
According to the offloading policy, the monitoring data is split into two parts: one is offloaded to the MEC server for processing, and the remaining part is kept locally for processing by the intelligent terminal.
The intelligent fault diagnosis system based on the MEC framework can be divided into three models: the network model, the communication model, and the computing model, which will be introduced separately in the following.

\subsection{Network model of intelligent fault diagnosis system}
The network of intelligent fault diagnosis system supporting offloading contains $M$ MEC servers and $K$ intelligent terminals, as shown in Fig. 1.
Let $\mathcal{M}=\{1,\cdots,M\} $ and $\mathcal{K}=\{1,\cdots,K\}$ be the index sets of the MEC servers and the intelligent terminals, respectively. 
Part of the diagnostic data will be offloaded to the MEC server, assuming that the MEC server has more computing power than the intelligent terminal.
The system time is divided into consecutive time frames with equal time period $\tau_0$ and the time indexed by $t\in T=\{0,1,\cdots\}$. 
The channel state information between the $m$-th MEC server and the $k$-th intelligent terminal is denoted as $h_{m,k}$, and the task size at intelligent terminal $k$ is marked as $C_k$. 
The channel state information of the MEC network $\{h_{m,k}(t)\}$ and the task arrival $C_k(t)$ at each intelligent terminal change for each time interval $t\in T$. 
In order to save the energy consumption of intelligent terminals and MEC servers and reduce the task processing latency, the central agent node needs to determine the task ratio of local execution content size and offloading content size, as well as the power allocation ratio of local task processing and data transmission.
The power splitting of the MEC server among multiple smart terminals should be determined if one MEC server is selected to help handle tasks from multiple intelligent terminals.
The communication model and the computational model are described in detail below.

\subsection{Communication model of MEC servers and intelligent terminals}
In the considered network of intelligent fault diagnosis systems, the communications are operated in an orthogonal frequency division multiple access framework, and a dedicated subchannel with bandwidth $B$ is allocated for each intelligent terminal for the partial task offloading. Supposing that intelligent terminal $k$ communicates with MEC server $m$, the received signal at MEC $m$ receiver can be represented as
\begin{equation}\label{rui_1}\nonumber
y_{m,k} = h_{m,k} \sqrt{p_k^o(t)} s_k + n_{m,k},
\end{equation}
where $s_k $ denotes the symbols transmitted from intelligent terminal $k$, $p_k^o(t)$ is the utilized power at intelligent terminal $k$, and $n_{m,k}$ denotes the received additive Gaussian noise with power $N_0$. Here the channel gains $h_{m,k}(t)$ follows the finite-state Markov chain (FSMC), and thus the communication rate between MEC server $m$ and intelligent terminal $k$ is give by
\begin{equation}\label{rui_2}
\begin{aligned}
r_{m,k}^o(t)=B\log_2\left(1+\frac{p_k^o(t) |h_{m,k}(t)|^2}{N_0}\right).
\end{aligned}
\end{equation}

\subsection{Computing model of intelligent fault diagnosis system}
The task $C_k(t)$ received at intelligent terminal $k$ at time $t$ need to be processed during time interval $t$. Denote the task splitting ratio as $\alpha_k\in [0,1]$ which indicates that at time interval $t$, $\alpha_kC_k(t)$ bits are executed at the intelligent terminal device and the remaining $(1-\alpha_k)C_k(t)$ bits are offloaded to and processed by the MEC server.

1) \textit{Local computing:} 
In local computation, the CPU of the intelligent terminal device is the primary engine, which adopts the dynamic frequency and voltage scaling (DVFS) technique and the performance of the CPU is controlled by the CPU-cycle frequency $\kappa_u$. 
Let $p_k^l(t)$ denote the local processing power at intelligent terminal $k$, then the intelligent terminal's computing speed (cycles per second) $f_k^l(t)$ at $t$-th slot is given by
\begin{equation}\nonumber
\begin{aligned}
f_k^l(t)=\sqrt[3]{\frac{p_k^l(t)}{\kappa_u}}.
\end{aligned}
\end{equation}

Let $D_k$ {denote} the number of CPU cycles required for intelligent terminal $k$ to accomplish one task bit. 
Then the local computation rate for intelligent terminal $k$ at time slot $t$ is given by
\begin{equation}\label{comp_rate_intelligent terminal}
\begin{aligned}
r_k^l=\frac{f_k^l}{D_k}=\frac{\sqrt[3]{\frac{p_k^l}{\kappa_u}}}{D_k}.
\end{aligned}
\end{equation}

2) \textit{Mobile Edge Computation Offloading:} 
The task model for mobile edge computation offloading is the data-partition model, where the task-input bits are bit-wise and can be arbitrarily divided into different groups. 
At the beginning of the time slot, the intelligent terminal chooses which MEC server to connect to according to the channel state.
Assume that the processed power which is allocated to the intelligent terminal $k$ by the MEC server $m$ is $p_{m,k}^c$, then the computation rate $r_{m,k}^c$ at MEC server $m$ for intelligent terminal $k$ is:
\begin{equation}\label{comp_rate_MEC}
\begin{aligned}
r_{m,k}^c=\frac{\sqrt[3]{\frac{p_{m,k}^c}{\kappa_m}}}{D_m},
\end{aligned}
\end{equation}
where $D_m$ is the number of CPU cycles required for the MEC server to accomplish one task bit, and $\kappa_m$ denotes the CPU-cycle frequency at the MEC server. 
It is noted that the MEC server can simultaneously process tasks from multiple intelligent terminals.
We assume multiple applications can be executed parallel with a negligible processing latency. 
The feedback time from the MEC to the intelligent terminal is ignored due to the small sized computational output.

\section{DQN-Based Offloading Design}
In this section, we develop a DQN-based offloading framework for minimizing the long-term processing delay cost. 
With the development of the traditional Q-learning algorithm, DQN is particularly suitable for high-dimensional state spaces and possesses fast convergence behavior. 
The MEC system constructs the DQN environment in the considered DQN offloading design framework. 
A central agent node is set up to observe status, perform actions and receive feedback rewards.
The center can be the cloud server or a MEC server.

The DQN-based offloading framework is introduced in the following, in which the corresponding state space, action space, and reward are defined. 
In the overall DQN paradigm, it is assumed that the instantaneous CSI is estimated at MEC servers using the training sequences and then delivered to the agent. 
The CSI observed at the agent is the delayed version due to the channel estimation operations and feedback delay. 
Only local CSI of intelligent terminals which connect to this MEC server is acquired for each MEC server.

\subsection{System state and action spaces}
{\bf System State Space}:
In the considered DQN paradigm, the state space observed by the agent includes the CSI of the overall network and the received task size $C_k(t)$ at time $t$. 
As the agent needs to consume extra communication overhead to connect the CSI from all MEC servers, the MEC server at time $t$ observes a delayed version of CSI at time $t-1$, i.e., $\{h_{m,k}(t-1)\}$.  
Denote
\begin{equation} \label{state}
\begin{aligned}
{\bf h}(t) &= \big\{ h_{1,1}(t), h_{1,2}(t),\cdots, h_{M,K}(t) \big\}, \\
{\bf C}(t) &= \big\{ C_1(t), C_2(t),\cdots, C_K(t) \big\}.
\end{aligned}
\end{equation}

The state space observed at time $t$ can be represented as
\begin{equation}\nonumber
\begin{aligned}
S(t)= \big\{ {\bf h}(t-1), {\bf C}(t) \big\}.
\end{aligned}
\end{equation}

{\bf System Action Space}:
The agent will take certain actions to interact with the environment with the observed state space $S(t)$. 
As DQN can only take care of the discrete actions, the actions defined in the proposed DQN paradigm constitute only the MEC server selection.
%
%
The MEC server selection action is denoted  as $a(t)$, which can be represented as
\begin{equation}\nonumber
\begin{aligned}
a(t) = \big\{x_{m,k}(t)|x_{m,k}(t)\in \{0,1\}\big\},
\end{aligned}
\end{equation}
where $x_{m,k}(t)=0$ means that the intelligent terminal $k$ does not select the MEC server $m$ at $t$-th time slot, while $x_{m,k}(t)=1$ indicates that the intelligent terminal $k$ selects the MEC server $m$ at $t$-th time slot.

\subsection{Reward Function}

In the DQN paradigm, the reward is defined as the maximum time delay required to complete all the tasks received at all intelligent terminals. 
After taking the actions, a dedicated MEC server can calculate the time delays required for the intelligent terminals choosing this MEC server to offload, as all MEC can observe the local CSI. 
With the loss of generality, we assume that intelligent terminal $k$ with $k\in \mathcal{O}_k$ offloads the tasks to MEC $m$, where set $\mathcal{O}_k$ defines the indexes of the intelligent terminals selecting MEC server $m$ to offload tasks.  
To minimize the required time delays, the MEC server needs to formulate an optimize problem to find optimal $\alpha_k(t)$, $p_k^l(t)$, $p_k^o(t)$, and $p_{m,k}^c(t)$. 
It is worth noting that as the MEC server knows the instantaneous CSI at time $t$, the solution can be obtained based on ${\bf h}(t)$, which is different from the MEC server selection taken based on ${\bf h}(t-1)$. 
For the intelligent terminals which do not offload tasks to the MEC servers, the required time delays for local task processing can be known by these intelligent terminals. 
The agent collects all the time delay consumptions from the intelligent terminals and the MEC servers to obtain the final reward.



We detail how to compute the time delay for intelligent terminal $k$, assuming that it selects MEC server $m$ to offload. 
The total time consumption for completing the task processing at intelligent terminal $k$ is denoted as $t_k$, which equals to $t_k=\max\{t_k^l, t_{m,k}^o+t_{m,k}^c\}$ where $t_k^l$, $t_{m,k}^o$, and $t_{m,k}^c$ denote the times required for intelligent terminal local task processing, task offloading transmission from intelligent terminal $k$ to MEC server $m$, and task processing at MEC server, respectively.

With the computation rate $r_k^l$ defined in eq. \eqref{comp_rate_intelligent terminal}, time $t_k^l$ can be represented as
\begin{equation}\nonumber
t_k^l= \frac{\alpha_k C_k(t)}{r_k^l (t)}.
\end{equation}

As the size of the offloaded task is $(1-\alpha_k)C_k$ with the communication rate defined in {eq. \eqref{rui_2}}, time $t_{m,k}^o$ can be calculated as
%
\begin{equation} \nonumber
t_{m,k}^o=\frac{(1-\alpha_k)C_k(t)}{r_{m,k}^o(t)}.
\end{equation}

With the computation rate $r_{m,k}^c$ allocated by MEC server $m$ to intelligent terminal $k$ in eq. \eqref{comp_rate_intelligent terminal}, time $t_{m,k}^c$ can be computed as
\begin{equation}\nonumber
t_{m,k}^c=\frac{(1-\alpha_k)C_k(t)}{r_{m,k}^c(t)}.
\end{equation}

To maximize the reward, we need to minimize the time delay for each intelligent terminal under the total energy constraint at intelligent terminals and MEC servers. 
To illustrate the way to find optimal $\alpha_k(t)$, $p_k^l(t)$, $p_k^o(t)$, and $p_{m,k}^c(t)$ for different types of MEC server selection, we next present two typical offloading scenarios, that is, a MEC server serves one intelligent terminal and a MEC server serves two intelligent terminals. 
It is noted that the proposed way of solving $\alpha_k(t)$, $p_k^l(t)$, $p_k^o(t)$, and $p_{m,k}^c(t)$ can be extended to the case where a MEC server serves arbitrary number of intelligent terminals.


1) \textit{Scenario 1: one MEC server serves one intelligent terminal}

The energy consumption at intelligent terminal $k$, denoted by $E_k$, includes two parts, i.e., one part for local partial task processing and another for partial task transmission. Therefore, $E_k$ can be written as
\begin{equation}\nonumber
\begin{aligned}
E_k=p_k^l(t) t_k^l+p_k^o(t) t_{m,k}^o.
\end{aligned}
\end{equation}

The energy consumption at the MEC server $m$ for processing the partial task offloaded from intelligent terminal $k$ {is} denoted by $E_{m, k}$, and can be represented as
\begin{equation}\nonumber
\begin{aligned}
E_{m, k}=p_{m,k}^c(t) t_{m,k}^c.
\end{aligned}
\end{equation}

The optimization problem formulated to find optimal $x_k(t)=\{\alpha_k(t), p_k^l(t), p_k^o(t), p_{m,k}^c(t)\}$ is given by
\begin{subequations}\label{eqn-lp}
	\begin{eqnarray}
	\min_{x_k(t) }  && \ t_k=\max\{t_k^l, t_{m,k}^o+t_{m,k}^c\} \\
	{\rm s.t.} && E_k \leq E_{{\rm max}, k},  \forall k \in \mathcal{K}   \\
	&& E_m \leq E_{{\rm max},m},  \forall m \in \mathcal{M},
	\end{eqnarray}
\end{subequations}
where $E_{{\rm max}, k}$ and $E_{{\rm max},m}$ denote the maximum available energy at intelligent terminal $k$ and MEC server $m$, respectively.
Problem \eqref{eqn-lp} can be rewritten as
\begin{subequations}\label{eqn_2}
	\begin{eqnarray}
	\min_{x_k(t)} &&\max \bigg\{\frac{{\alpha}_ k(t) C_k(t)}{\frac{\sqrt[3]{\frac{ p_1^k(t)}{\kappa_l}}}{D_k}},  \frac{(1-{\alpha_k(t)})C_k(t)}{\frac{\sqrt[3]{\frac{ p_{m,k}^c(t)}{\kappa_m}}}{D_m} } \label{eqn_2_a}\\
	&&  +  \frac{(1-{\alpha_k(t)})C_k(t)}{ B\log_2(1+\frac{ p_k^o(t)|h_{m,k}(t)|^2}{N_0})} \bigg\} \nonumber\\
	{\rm s.t.} &&  p_k^l (t) t_k^l+ p_k^o (t) t_{m,k}^o \leq E_{{\rm max}, k} \label{eqn_2_b} \\
	&& p_{m,k}^c (t) t_{m,k}^c \leq E_{{\rm max},m}  \label{eqn_2_c}\\
	&& {\alpha}_k \in [0,1]. \label{eqn_2_d}
	\end{eqnarray}
\end{subequations}

To solve problem \eqref{eqn_2}, we first find that at optimal solution, constraint \eqref{eqn_2_c} must be active, which can minimize the the objective value \eqref{eqn_2_a}. We thus have
\begin{equation}\label{eqn_2_1}\nonumber
\begin{split}
p_{m,k}^c(t) t_{m,k}^c = p_{m,k}^c(t) \frac{(1-{\alpha_k(t)})C_k(t)}{\frac{\sqrt[3]{\frac{ p_{m,k}^c(t)}{\kappa_m}}}{D_m} }=E_{{\rm max},m},
\end{split}
\end{equation}
which produces
\begin{equation}\label{eqn_2_2}
\begin{split}
p_{m,k}^c(t)  =  \left(\frac{ E_{{\rm max},m}  }{(1-{\alpha_k(t)})C_k(t) \kappa_m^{1/3} D_m}\right)^{3/2}.
\end{split}
\end{equation}

Substituting \eqref{eqn_2_2} to problem \eqref{eqn_2}, we have
\begin{subequations}\label{eqn_2_3}
	\begin{eqnarray}
	\min_{x_k(t)} &&\max \bigg\{\frac{{\alpha}_k(t) C_k(t)}{(\frac{\sqrt[3]{\frac{ p_k^l(t)}{\kappa_l}}}{D_l})},  \frac{ (1-{\alpha_k(t)})^{3/2} C_k(t)^{3/2} } { \kappa_m^{ 1/6} D_m^{1/2} E_{{\rm max},m}^{1/2}} \\
	&&  + \frac{(1-{\alpha_k(t)})C_k(t)}{ B\log_2(1+\frac{ p_k^o(t) |h_{m,k}|^2}{N_0})} \bigg\} \label{eqn_2_3_a} \nonumber\\
	{\rm s.t.}\quad &&  p_k^l(t) t_k^l+ p_k^o(t) t_{m,k}^o \leq E_{{\rm max}, k} \label{eqn_2_3_b} \\
	&& {\alpha}_k \in [0,1]. \label{eqn_2_3_d}
	\end{eqnarray}
\end{subequations}

It is noted that problem \eqref{eqn_2_3} is a non-convex optimization problem. 
We propose an alternating algorithm to solve $p_k^l(t)$ and $p_k^o(t)$ and ${\alpha}_k(t)$ in different subproblems separately to find an efficient solution.
In the first subproblem, we solve $p_k^l(t)$ for given ${\alpha}_k(t)$, and $p_k^o(t)$. 
To minimize the objective function, the optimal solution of $p_k^l(t)$ should activate constraint \eqref{eqn_2_3_b}, that is, $p_k^l(t) t_k^l = E_{{\rm max}, k}- p_k^o(t) t_{m,k}^o$, which implies
\begin{equation}\label{eqn_2_4}
\begin{split}
p_k^l(t) = \left( \frac{E_{{\rm max}, k}- p_k^o(t) t_{m,k}^o}{ {\alpha}_k(t) C_k(t) D_l \kappa_l^{1/3} } \right)^{3/2}.
\end{split}
\end{equation}

In the second subproblem, we solve $p_k^o(t)$ with given $p_k^l(t)$ and ${\alpha}_k(t)$. 
The corresponding optimization problem is given by
\begin{subequations}\label{eqn_2_44}
	\begin{eqnarray}
	\min_{p_k^o(t)} && \frac{(1-{\alpha_k(t)})C_k(t)}{ B\log_2(1+\frac{ p_k^o(t) |h_{m,k}|^2}{N_0})}  \label{eqn_2_4_a}\\
	{\rm s.t.}  &&    \frac{(1-{\alpha_k(t)})C_k(t) p_k^o(t)}{ B\log_2(1+\frac{ p_k^o(t) |h_{m,k}|^2}{N_0})} \leq E_{{\rm max}, k} -p_k^l(t) t_k^l. \label{eqn_2_4_b}
	\end{eqnarray}
\end{subequations}
Problem \eqref{eqn_2_4} is a convex optimization problem and can be efficiently solved, such as the interior point algorithm, etc.

In the third problem, ${\alpha}_k(t)$ is solved with given $p_k^l(t)$ and $p_k^o(t)$. 
The corresponding optimization problem is given by
\begin{subequations}\label{eqn_2_5}
	\begin{eqnarray}
	\min_{{\alpha}_k(t)} &&\max \bigg\{\frac{{\alpha}_k(t) C_k(t) }{(\frac{\sqrt[3]{\frac{ p_k^l(t)}{\kappa_l}}}{D_l}},  \frac{ (1-{\alpha_k(t)})^{3/2} C_k(t)^{3/2} } { \kappa_m^{ 1/6} D_m^{1/2} {E_{{\rm max},m}}^{1/2}}  ~~~\\
	&& +\frac{(1-{\alpha_k(t)})C_1}{ B\log_2(1+\frac{ p_k^o(t) |h_{m,k}|^2}{N_0})} \bigg\} \label{eqn_2_5_a}\nonumber \\
	{\rm s.t.} && {\alpha}_k(t) \in [0,1]. \label{eqn_2_5d}
	\end{eqnarray}
\end{subequations}
For the min-max problem \eqref{eqn_2_5}, by denoting $f_1(\alpha_k(t)) = \frac{ \alpha_k(t) C_k(t)}{\frac{\sqrt[3]{\frac{ p_k^l(t)}{\kappa_l}}}{D_l}}$, $f_2(\alpha_k(t)) =\frac{ (1-{\alpha_k(t)})^{3/2} C_k(t)^{3/2} } { \kappa_m^{ 1/6} D_m^{1/2} {E_{{\rm max},m}}^{1/2}} + \frac{(1-{\alpha_k(t)})C_k(t)}{ B\log_2(1+\frac{ p_k^o(t) |h_{m,k}|^2}{N_0})}$ and $f(\alpha_k(t))=f_1(\alpha_k(t))+f_2(\alpha_k(t))$, it is known that the optimal $\alpha_k(t)$, denoted by $\alpha^*_k(t)$, occurs in the following three cases, that is, $\alpha^{\rm 1}_k(t)=0$, $\alpha^{\rm 2}_k(t)=1$ or $f_1(\alpha^{\rm 3}_k(t))=f_2(\alpha^{\rm 3}_k(t))$. Note that the solution of the third case can be obtained by solving a cubic equation. The final solution is given as
\begin{equation}\label{eqn_2_6}
\begin{split}
\alpha^*_k(t) = \left \{
\begin{array}{cc}
\alpha^{\rm 1}_k(t) & {\rm if}~~ f(\alpha^{\rm 1}_k(t))\leq \{f(\alpha^{\rm 2}_k(t)),f(\alpha^{\rm 3}_k(t)) \} \\
\alpha^{\rm 2}_k(t) & {\rm if}~~ f(\alpha^{\rm 2}_k(t))\leq \{f(\alpha^{\rm 1}_k(t)),f(\alpha^{\rm 3}_k(t)) \} \\
\alpha^{\rm 3}_k(t) & {\rm if}~~ f(\alpha^{\rm 3}_k(t))\leq \{f(\alpha^{\rm 1}_k(t)),f(\alpha^{\rm 2}_k(t)) \} \\
\end{array}
\right. .
\end{split}
\end{equation}

By alternating three subproblems with the solutions given in \eqref{eqn_2_4}, \eqref{eqn_2_44} and \eqref{eqn_2_6} until convergence, we obtain the final solution.


2) \textit{Scenario 2: one MEC server serves two intelligent terminals}
%

Assume that MEC server $m$ serves two intelligent terminals, {e.g.,} intelligent terminal $k$ and intelligent terminal $k^\prime$, then the optimization problem can be formulated as follows
\begin{subequations}\label{eqn_2_66}\nonumber
	\begin{eqnarray}
	\min  && \max \big\{ t_k^l,  t_{m,k}^o + t_{m,k}^c, t_{k^\prime}^l,  t_{m,k^\prime}^o + t_{m,k^\prime}^c \big\} \nonumber\\
	{\rm s.t.}  &&  p_k^l(t) t_k^l+ p_{k}^o(t) t_{m,k}^o \leq E_{{\rm max}, k} \nonumber\\
	&&  p_{k^\prime}^l(t) t_{k^\prime}^l+ p_{{k^\prime}}^o(t) t_{m,{k^\prime}}^o \leq E_{{\rm max}, {k^\prime}} \nonumber\\
	&&  p_{m,k}^c(t) t_{m,k}^c + p_{m,k^\prime}^c(t) t_{m,k^\prime}^c \leq E_{{\rm max},m}. \nonumber
	\end{eqnarray}
\end{subequations}

The previously proposed iterative algorithm can still be applied here to solve $\alpha_i(t)$, $p^l_i(t)$, $p^o_i(t)$ and $p_{m,i}^c(t)$ with $i=\{k, k^\prime\}$. Here the only difference lies in solving $p^c_{m,k}(t)$ and $p^c_{m,k^\prime}(t)$. The corresponding optimization problem can be formulated as
\begin{subequations}\label{equ_new_2}  \nonumber
	\begin{eqnarray}
	\min &&\max\bigg\{t_{m,k}^o+ \frac{(1-{\alpha_k(t)})C_k(t)}{\frac{\sqrt[3]{\frac{ p_{m,k}^c(t)}{\kappa_m}}}{D_m} },  \nonumber\\
	&& t_{m,k^\prime}^o+ \frac{(1 - \alpha_{k^\prime}(t))C_{k^\prime}(t)}{\frac{\sqrt[3]{\frac{ p_{m,k^\prime}^c(t)}{\kappa_m}}}{D_m} } \bigg\}  \nonumber \\
	\mbox{s.t.} &&  \frac{(1-{\alpha_k(t) })C_k(t)}{\frac{\sqrt[3]{\frac{ p_{m,k}^c(t) }{\kappa_m}}}{D_m} } + \frac{(1-\alpha_{k^\prime}(t))C_{k^\prime}(t)}{\frac{\sqrt[3]{\frac{ p_{m,k^\prime}^c(t)}{\kappa_m}}}{D_m} } \leq E_{{\rm max},m}.  \nonumber
	\end{eqnarray}
\end{subequations}

It is worth noting that the optimal solution must activate the constraints and make the two terms within the objective function equal to each other. 
Therefore, the optimal $p^c_{m,k}(t)$ and $p^c_{m,k^\prime}(t)$ can be obtained by solving the following equations
\begin{equation}\nonumber
\begin{aligned}
t_{m,k}^o+ \frac{(1-{\alpha_k(t)})C_k(t)}{\frac{\sqrt[3]{\frac{ p_{m,k}^c(t)}{\kappa_m}}}{D_m} } = t_{m,k^\prime}^o+ \frac{(1 - \alpha_{k^\prime}(t))C_{k^\prime}(t)}{\frac{\sqrt[3]{\frac{ p_{m,k^\prime}^c(t)}{\kappa_m}}}{D_m} } \\
\frac{(1-{\alpha_k(t) })C_k(t)}{\frac{\sqrt[3]{\frac{ p_{m,k}^c(t) }{\kappa_m}}}{D_m} } + \frac{(1-\alpha_{k^\prime}(t))C_{k^\prime}(t)}{\frac{\sqrt[3]{\frac{ p_{m,k^\prime}^c(t)}{\kappa_m}}}{D_m} } = E_{{\rm max},m}.
\end{aligned}
\end{equation}

\begin{figure}[t]
	\centerline{\includegraphics[width=0.8\textwidth]{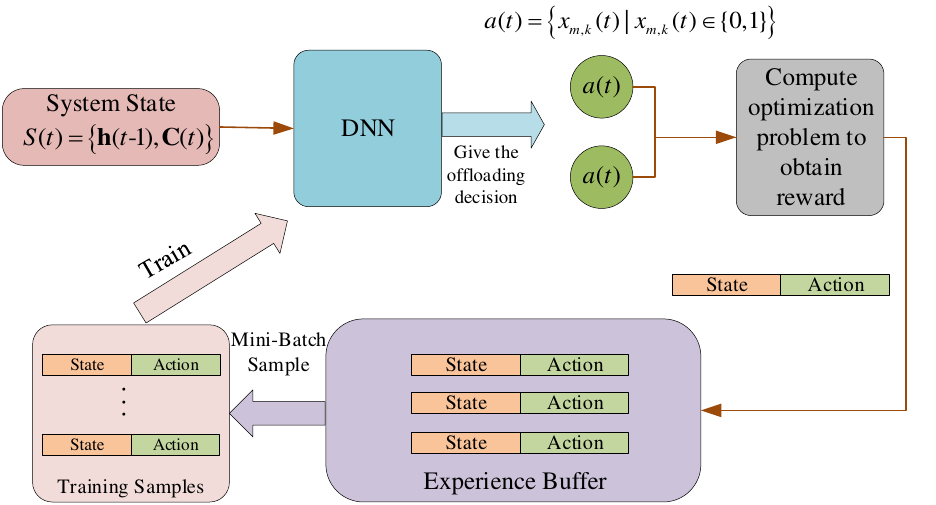}}
	\caption{The structure of the DQN-based offloading algorithm.}
	\label{DQN}
\end{figure}

Hence, under an action $a(t)$, the system reward can be obtained as
\begin{equation}\label{reward}
r_t =- \max \big\{t_k| k\in \mathcal{K}, a(t) \big\}.
\end{equation}
The structure of the DQN-based offloading algorithm is illustrated in Fig. \ref{DQN}, and the pseudocode is presented in Algorithm 1.

\begin{algorithm}[h]
	\caption{The DQN-based Offloading Algorithm}
	\begin{algorithmic}[1]
		\STATE Initialize the experience replay buffer $B$;\
		\STATE Initialize action-value function $Q$ with random weights $\theta$;\
		\STATE Initialize target action-value function $Q'$ with random weights $\theta^-=\theta$;\
		\FOR{each episode $n=1,2,\cdots,N$}
		\STATE Reset simulation parameters for the environment;\
		\STATE Randomly generate an initial state $s_1$;\
		\FOR{each time slot $t=1,2,...,T$}
		\STATE Generate an action $a_t=\mu(s_t|\theta^\mu)+\nabla\mu$ to determine which MEC server to connect to;\
		\STATE Execute action $a_t$ and solving corresponding optimization to obtain reward $r_t$;\
		\STATE Receive reward $r_t$ and observe the next state $s_{t+1}$;\
		\STATE Store the tuple $(s_t,a_t,r_t,s_{t+1})$ into $B$;\
		\STATE Sample a random mini-batch of $N$ transitions $(s_t,a_t,r_t,s_{t+1})$ from $B$;\
		\STATE Perform gradient descent and update Q-network;\
		\STATE Every $C$ steps reset $Q'=Q$;\
		\ENDFOR
		\ENDFOR
	\end{algorithmic}
\end{algorithm}

\section{DDPG-Based Offloading Design}

Note that only the discrete actions can be handled by the DQN-based offloading design, where the reward acquisition mainly depends on solving the formulated optimization problems at MEC servers, which may increase the extra computing burden at the MEC servers. 
In this section, we rely on the DDPG to design offloading policy, considering that DDPG can deal with discrete and continuous value actions. 
Different from DQN, DDPG uses the Actor-Critic network to improve the accuracy of the model. 
In this section, we directly regard $a(t)$, $\alpha_k(t)$, $p_k^l(t)$, $p_k^o(t)$, and $p_{m,k}^c(t)$ as the output action instead of disassembling the problem into two parts.

{\bf System State Space}: 
In the DDPG offloading paradigm, the system state space action is the same as the DQN-based offloading paradigm, which is given by
\begin{equation}\nonumber
\begin{aligned}
S(t)= \big\{ {\bf h}(t-1), {\bf C}(t) \big\},
\end{aligned}
\end{equation}
where ${\bf h}(t-1)$ and ${\bf C}(t)$ defined in \eqref{state}. 
As in the DQN offloading paradigm, the agent can only observe the delayed version of CSI due to channel estimation operations and feedback delay.

{\bf System Action Space}:
In DDPG offloading paradigm, the value of $p_{m,k}^c(t)$ is utilized to indicate the MEC server selection, which $p_{m,k}^c(t)=0$ represents that there is no partial task at intelligent terminal $k$ offloaded to the MEC server $m$. 
In other words, the MEC server $m$ is not chosen by intelligent terminal $k$. 
If $p_{m,k}^c(t)$ is not equal to $0$, it means that the intelligent terminal $k$ decides to offload partial tasks to the MEC server $m$. 
Since the intelligent terminal can only connect to one MEC server at one time slot, only one $p_{m,k}^c(t)$ in any time slot is not 0, and the remaining ones are $0$. 
The action space of DDPG offloading paradigm can be expressed as
\begin{equation}\nonumber
a(t) = \big\{\alpha_k(t), p_k^l(t), p_k^o(t), p_{m,k}^c(t) \big\}, \forall k, m.
\end{equation}
It is noted that here the continuous actions $p_k^l(t), p_k^o(t), p_{m,k}^c(t)$ can be obtained based on state $S(t)$ with delayed CSI ${\bf h}(t-1)$.

\begin{figure}[t]
	\centerline{\includegraphics[width=0.85\textwidth]{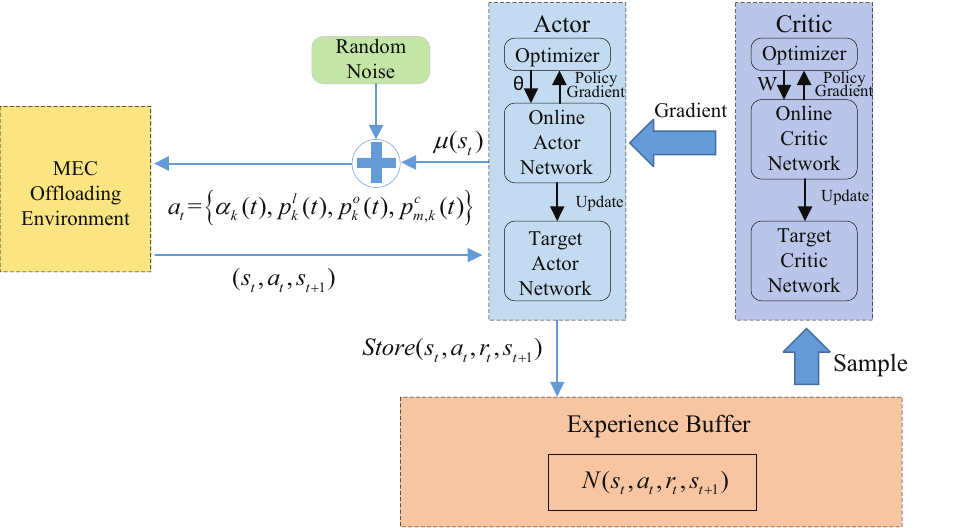}}
	\caption{The structure of the DDPG-based offloading algorithm.}
	\label{DDPG}
\end{figure}
{\bf System Reward Funciton}:
In the DDPG offloading algorithm, $\alpha_k(t)$, $p_k^l(t)$, $p_k^o(t)$, and $p_{m,k}^c(t)$ can be obtained from a continuous action space. 
With the decisions, the agent tells each intelligent terminal $k$ the selected MEC server and delivers $p_k^l(t)$, $p_k^o(t)$ to it to perform the offloading. 
Moreover, the agent needs to send $p_{m,k}^c(t)$ to each server to allocate computing resources. 
After that, the reward is obtained as in \eqref{reward} by collecting $t_k$ observed at the MEC servers or intelligent terminals.

Compared to the DQN-based offloading paradigm, the DDPG-based offloading paradigm does not need the MEC servers to solve the optimization problems, which can release the computation burden at the MEC servers. 
However, as the DDPG algorithm is generally more complex than the DQN algorithm, the computation complexity unavoidably increases at the agent.
The structure of the DDPG-based offloading algorithm is illustrated in Fig.~\ref{DDPG}. 
We provide the pseudocode in Algorithm 2.

\begin{algorithm}[t]
	\caption{The DDPG-based Offloading Algorithm}
	\begin{algorithmic}[1]
		\STATE Randomly initialize the actor network $\mu_{\theta^\mu}$ and the critic network $Q_{\theta^Q}$ with weights $\theta^\mu$ and $\theta^Q$;\
		\STATE Initialize target network $\mu$ and $Q$ with weights $\theta^{\mu'}\leftarrow\theta^\mu$, $\theta^{Q'}\leftarrow\theta^Q$;\
		\STATE Initialize the experience replay buffer $B$;\
		\FOR{each episode $n=1,2,\cdots,N$}
		\STATE Reset simulation parameters for the environment;\
		\STATE Randomly generate an initial state $s_1$;\
		\FOR{each time slot $t=1,2,...,T$}
		\STATE Select an action $a_t=\mu(s_t|\theta^\mu)+\nabla\mu$ to determine the power for transmission and computation;\
		\STATE Execute action $a_t$ and receive reward $r_t$ and observe the next state $s_{t+1}$;\
		\STATE Store the tuple $(s_t,a_t,r_t,s_{t+1})$ into $B$;\
		\STATE Sample a random mini-batch of $N$ transitions $(s_t,a_t,r_t,s_{t+1})$ from $B$;\
		\STATE Update the critic network by minimizing the loss $L:$\\ $L=\frac{1}{N}\sum_{t=1}^{N}\big(r_t+\max \limits_{a\in A}Q(s_t^{'},a|\theta^{Q'})-Q(s_t,a_t|\theta^Q)\big)^2;$
		\STATE Update the actor network by using the sampled policy gradient:\\
		$\nabla_{\theta^\mu}J\approx\frac{1}{N}\sum_{t=1}^{N}\nabla_\alpha Q(s_t,a|\theta^Q)|_{a=a_t}\nabla_{\theta^\mu}\mu(s_t|\theta^\mu);$
		\STATE Update the target networks by:\\
		$\theta^{\mu'}\leftarrow\tau\theta^\mu+(1-\tau)\theta^{\mu'};$\\
		$\theta^{Q'}\leftarrow\tau\theta^Q+(1-\tau)\theta^{Q'};$
		\ENDFOR
		\ENDFOR
	\end{algorithmic}
\end{algorithm}

\begin{figure}[!htb]
	\centerline{\includegraphics[width=0.75\textwidth]{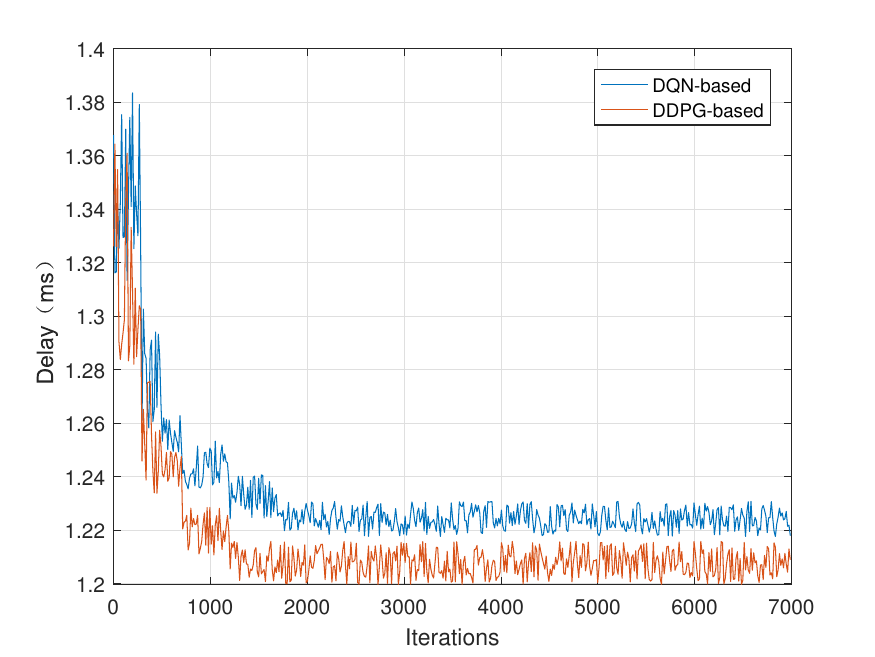}}
	\caption{The delay dynamics of each iteration of the DQN-based and DDPG-based algorithms during the training process, where the blue curve represents the delay dynamics of the DQN-based algorithm and the red curve represents the delay dynamics of the DDPG-based algorithm.}
	\label{fig5}
\end{figure}

\begin{figure}[!htb]
	\centerline{\includegraphics[width=0.7\textwidth]{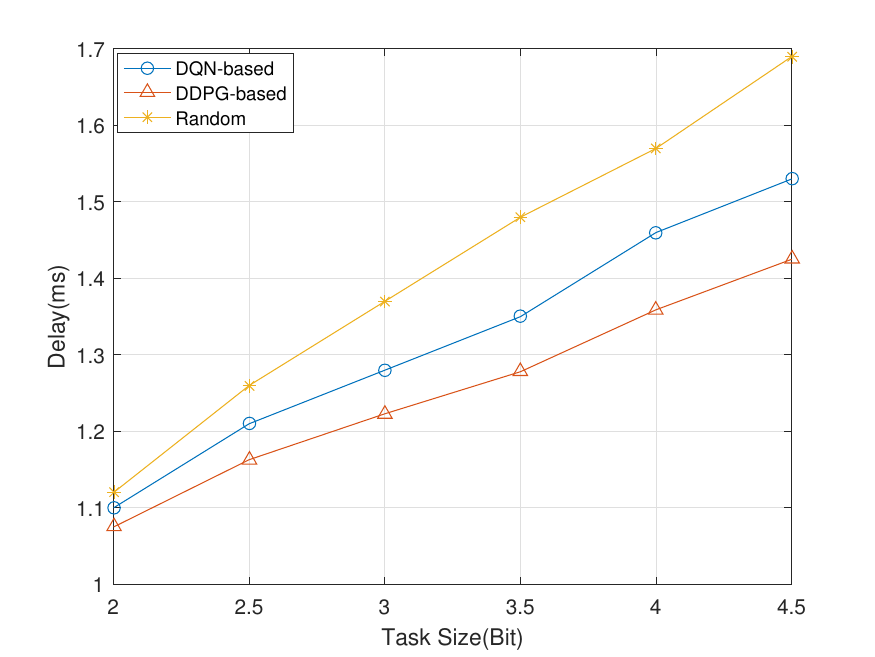}}
	\caption{The delay comparison under different task sizes, where the red curve represents the delay of the DDPG-based computational offloading paradigm, the blue curve represents the delay of the DQN-based computational offloading paradigm, and the yellow curve represents the delay of the "Random" policy.}
	\label{fig6}
\end{figure}

\begin{figure}[!htb]
	\centerline{\includegraphics[width=0.9\textwidth]{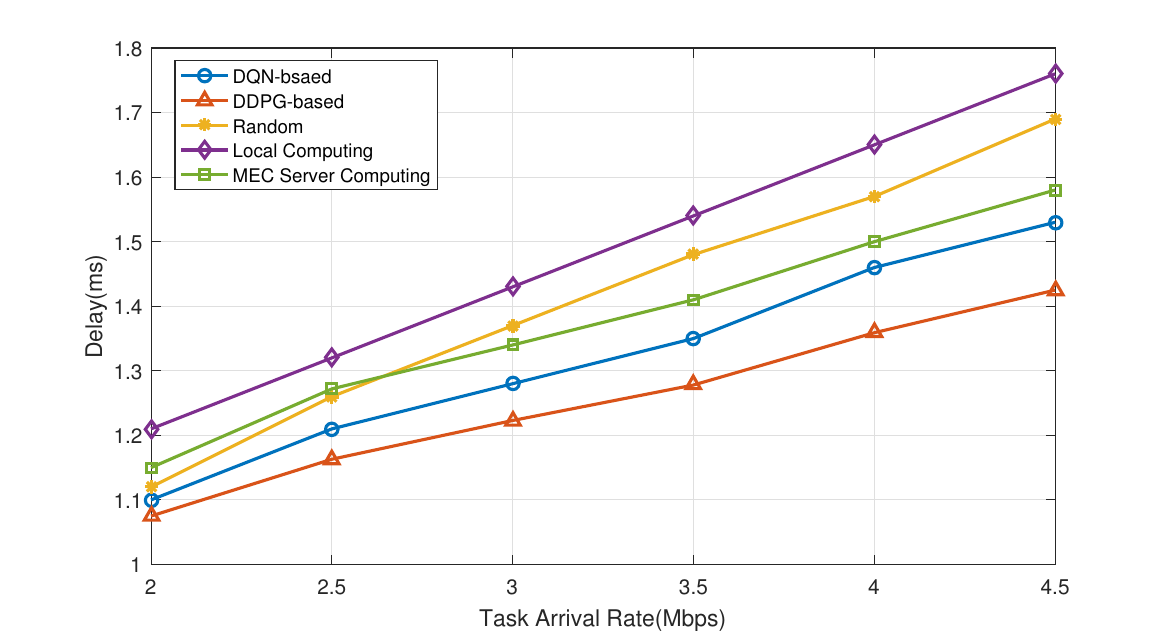}}
	\caption{The delay comparison under different task arrival rates, where the red curve represents the delay of the DDPG-based computational offloading paradigm, the blue curve represents the delay of the DQN-based computational offloading paradigm, the yellow curve represents the delay of the "Random" policy, the purple curve represents the delay of the "Local computing" policy, and the green curve represents the delay of the "MEC server computing" policy.}
	\label{fig7}
\end{figure}

\begin{figure}[!htb]
	\centerline{\includegraphics[width=0.9\textwidth]{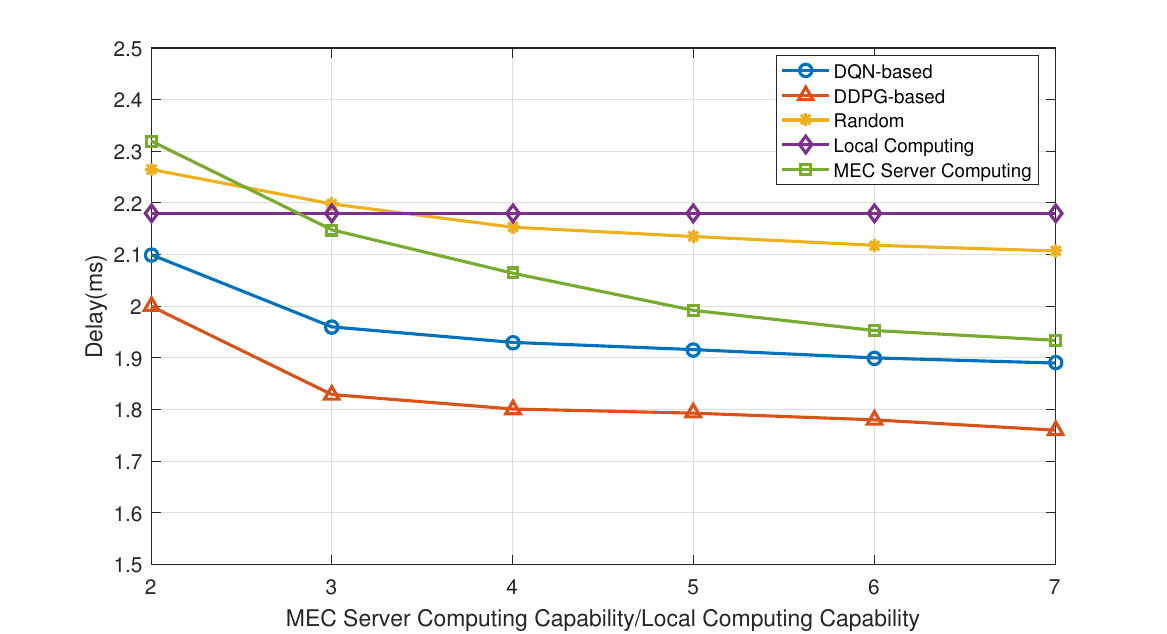}}
	\caption{The delay comparison under different computing capabilities, where the red curve represents the delay of the DDPG-based computational offloading paradigm, the blue curve represents the delay of the DQN-based computational offloading paradigm, the yellow curve represents the delay of the "Random" policy, the purple curve represents the delay of the "Local computing" policy, and the green curve represents the delay of the "MEC server computing" policy.}
	\label{fig8}
\end{figure}

\begin{figure}[!htb]
	\centerline{\includegraphics[width=0.9\textwidth]{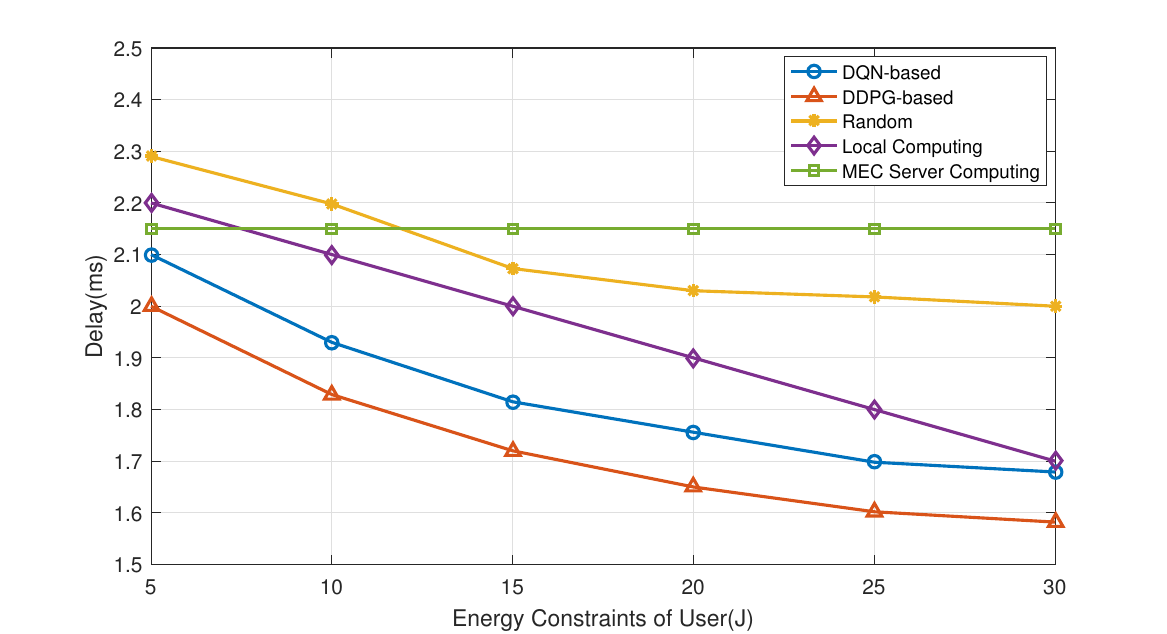}}
	\caption{The delay comparison under different energy constraints at the intelligent terminals, where the red curve represents the delay of the DDPG-based computational offloading paradigm, the blue curve represents the delay of the DQN-based computational offloading paradigm, the yellow curve represents the delay of the "Random" policy, the purple curve represents the delay of the "Local computing" policy, and the green curve represents the delay of the "MEC server computing" policy.}
	\label{fig9}
\end{figure}

\section{Numerical Results}\
In this section, we present the numerical simulation results to illustrate the performance of the proposed two offloading paradigms. 
Assume that the time interval of the system is $1$ ms, and the bandwidth of the intelligent fault diagnosis system is $1$ MHz.
Additionally, the required CPU cycles per bit are $300$ cycles/bit at the intelligent terminals and $120$ cycles/bit at MEC servers. 
In the training process, the learning rate of the DQN-based offloading algorithm is $0.01$. 
In the DDPG-based offloading algorithm, the learning rate of the actor network is $0.001$, and the learning rate of the critic network is $0.001$.
	
In Fig. \ref{fig5}, we plot the training process of the DQN-based algorithm and the DDPG-based algorithm, where the blue curve represents the delay dynamics of the DQN-based algorithm and the red curve represents the delay dynamics of the DDPG-based algorithm. 
The delay of the system is in an unstable state with large fluctuations in the beginning, indicating that the agent is constantly exploring the environment randomly. 
After a period of learning, the delay decreases slowly, and the fluctuation range gradually gets smaller. 
After about 1200 iterations, the DDPG-based algorithm converges to a stable value; after about 1500 iterations, the DQN-based algorithm converges. 
At this time, the average reward of each episode no longer changes, and the training process is completed. 
The DDPG-based algorithm converges faster and can obtain a lower latency than the DQN-based algorithm. 
This indicates that the performance of the DDPG-based algorithm is better than the DQN-based algorithm for our offloading problem.

In Fig. \ref{fig6}, the DDPG-based computational offloading paradigm, the DQN-based computational offloading paradigm, and the "Random" policy are compared for different task sizes in terms of delay.
"Random" means that the computing resources are allocated randomly.
The delay difference between the three policies is slight at task sizes below 2.5 Bit, with the DDPG-based computational offloading paradigm having the smallest delay and the "Random" policy having the largest delay.
The delay of the "Random" policy increases the most as the task size increases, while the latency of the DDPG-based computational offloading paradigm and the DQN-based computational offloading paradigm increases slightly less.
The delay of DDPG's computational offload paradigm and DQN-based computational offload paradigm consistently remains low compared to the "Random" policy.

In Fig. \ref{fig7}, we illustrate the offloading delay as the function of the amount of tasks at intelligent terminals. 
Three benchmarks, namely "Random", "Local computing", and "MEC server computing", are chosen to compare the performance with the proposed two offloading paradigms. 
Here "Random" means that the computing resources are allocated in a random manner; "Local computing" and "MEC server computing" mean that the tasks are processed only at intelligent terminals and only at MEC servers, respectively. 
The curves in Fig. \ref{fig7} show that the required time delay increases correspondingly as the amount of tasks grows. 
The computation delay of "Local computing" is the largest as intelligent terminals have little local computing capacity. 
"MEC server computing" performs better than "random scheme" when the task arrival rate is more significant than $2.7$ Mbps, which indicates that when the task arrival rate increases, task offloading to MEC servers can obtain a lower time delay. 
When the task arrival rate is greater than $4$ Mbps, the offloading time delay of "MEC server computing" is close to the DQN-based computation offloading algorithm, indicating that most tasks are offloaded to the MEC servers with large task sizes. 
Both proposed DQN and DDPG offloading paradigms achieve better performance than other benchmarks, proving the proposed methods' effectiveness. 
On the other hand, the DDPG-based computation offloading paradigm achieves a lower computation delay than the DQN-based computation offloading paradigm, which further verifies the superiority of the DDPG algorithm in dealing with high-dimensional continuous action-state space problems.

Fig. \ref{fig8} shows the impact of the computing capabilities of intelligent terminals and MEC servers on the processing delay. 
We fix the local computing capability as a constant value and increase the computing capacity of the MEC server continuously, so the computation delay of "Local computing" is not affected by the ratio of computing capacity between the intelligent terminal and MEC server. 
Under different computing capabilities, the proposed DQN and DDPG offloading paradigms can achieve better performance than the other three benchmarks, and the performance of the DDPG-based offloading paradigm is slightly better than the DQN-based offloading paradigm.
When the ratio of MEC server computing capacity to intelligent terminal computing capacity locates between 2 and 3 and the ratio increases, the processing speed of the MEC server is faster than the intelligent terminal, and the intelligent terminal chooses to offload more tasks to the MEC server. 
The computation delay of "MEC computing" is smaller than the "random scheme". 
When the ratio exceeds $3$ and as the ratio increases, the processing speed of the MEC server is significantly higher than the intelligent terminals. 
The intelligent terminal prioritizes the task offloading, and the task processing delay is still decreasing, but the downward trend slows down. 
The computation delay of "MEC computing" is lower than the "random scheme" and close to the DQN-based offloading paradigm, which indicates that most or all tasks are offloaded to the MEC servers. 
The decrease in the delay is mainly due to the increase in the computing capacity of the MEC servers.

Fig. \ref{fig9} illustrates the computation delay under different energy constraints at the intelligent terminals.
The curves show that "MEC computing" is not affected by the change in energy of the intelligent terminal.
"Local computing" highly depends on the intelligent terminal energy constraint, and the computation delay decreases significantly as the intelligent terminal energy increases. 
The increase in intelligent terminal energy indicates a fast local processing speed and high available transmission at the intelligent terminals, which can reduce the computation delay to a certain extent. 
The computation delay of DQN-based and DDPG-based offloading paradigms decreases significantly as the intelligent terminal's energy increases at the beginning.
The computation delay gradually decreases when the intelligent terminal's energy reaches a certain level, which shows that the intelligent terminal's energy constraint significantly impacts the computation delay within a specific range. 
The computation delay has a weaker impact when the intelligent terminal's energy exceeds a certain range. 
The DQN-based and the DDPG-based offloading paradigms achieve better performance than other offloading methods under different intelligent terminal energy constraints, which indicates the effectiveness of the proposed computational offloading algorithms. 
What's more, the performance of the DDPG-based offloading paradigm is slightly better than the DQN-based offloading paradigm.


\section{Conclusion}\
In this paper, we propose a novel framework for the intelligent mechanical fault diagnosis system, which is a resource allocation scheme based on deep reinforcement learning for offloading diagnostic data of multiple intelligent terminals.
The optimization parameters and optimization objectives can be determined by modeling the data offloading scenario of the intelligent fault diagnosis system.
Two deep reinforcement learning algorithms, i.e., DQN-based offloading strategy and DDPG-based offloading strategy, are investigated to solve the formulaic offloading optimization problem for obtaining the lowest latency. 
Comparing the different offloading schemes shows that the proposed deep reinforcement learning-based learning approach can reduce task processing latency under different system parameters.
The intelligent fault diagnosis framework proposed in this paper allows easier access to other intelligent technologies, such as deep learning techniques for data calibration, federated learning techniques and blockchain technologies for protecting user data privacy.

\section{Acknowledgements}
This work was supported by the National Science Foundation of China under Grant 62271352, the National Natural Science Foundation of China under Grant 12074254 and 51505277, the Natural Science Foundation of Shanghai under Grant 21ZR1434100, and Shanghai Science and Technology Innovation Action Plan Project No. 21220713100.
	
\numberwithin{equation}{section}

\appendix

\bibliographystyle{model1-num-names}
\bibliography{mybibfile}

\begin{thebibliography}{38}
\expandafter\ifx\csname natexlab\endcsname\relax\def\natexlab#1{#1}\fi
\providecommand{\url}[1]{\texttt{#1}}
\providecommand{\href}[2]{#2}
\providecommand{\path}[1]{#1}
\providecommand{\DOIprefix}{doi:}
\providecommand{\ArXivprefix}{arXiv:}
\providecommand{\URLprefix}{URL: }
\providecommand{\Pubmedprefix}{pmid:}
\providecommand{\doi}[1]{\href{http://dx.doi.org/#1}{\path{#1}}}
\providecommand{\Pubmed}[1]{\href{pmid:#1}{\path{#1}}}
\providecommand{\bibinfo}[2]{#2}
\ifx\xfnm\relax \def\xfnm[#1]{\unskip,\space#1}\fi
\bibitem[{Zhang et~al.(2019)Zhang, Chen, He, Yang, Gong, and
  Zhang}]{zhang2019privacy}
\bibinfo{author}{M.~Zhang}, \bibinfo{author}{J.~Chen}, \bibinfo{author}{S.~He},
  \bibinfo{author}{L.~Yang}, \bibinfo{author}{X.~Gong},
  \bibinfo{author}{J.~Zhang},
\newblock \bibinfo{title}{Privacy-preserving database assisted spectrum access
  for industrial internet of things: a distributed learning approach},
\newblock \bibinfo{journal}{IEEE Transactions on Industrial Electronics}
  \bibinfo{volume}{67} (\bibinfo{year}{2019}) \bibinfo{pages}{7094--7103}.
\bibitem[{Yang et~al.(2022)Yang, Xu, Lei, Lee, Stewart, and
  Roberts}]{yang2022multi}
\bibinfo{author}{B.~Yang}, \bibinfo{author}{S.~Xu}, \bibinfo{author}{Y.~Lei},
  \bibinfo{author}{C.-G. Lee}, \bibinfo{author}{E.~Stewart},
  \bibinfo{author}{C.~Roberts},
\newblock \bibinfo{title}{Multi-source transfer learning network to complement
  knowledge for intelligent diagnosis of machines with unseen faults},
\newblock \bibinfo{journal}{Mechanical Systems and Signal Processing}
  \bibinfo{volume}{162} (\bibinfo{year}{2022}) \bibinfo{pages}{108095}.
\bibitem[{Azamfar et~al.(2020)Azamfar, Li, and Lee}]{azamfar2020intelligent}
\bibinfo{author}{M.~Azamfar}, \bibinfo{author}{X.~Li},
  \bibinfo{author}{J.~Lee},
\newblock \bibinfo{title}{Intelligent ball screw fault diagnosis using a deep
  domain adaptation methodology},
\newblock \bibinfo{journal}{Mechanism and Machine Theory} \bibinfo{volume}{151}
  (\bibinfo{year}{2020}) \bibinfo{pages}{103932}.
\bibitem[{Aslam et~al.(2020)Aslam, Michaelides, and
  Herodotou}]{aslam2020internet}
\bibinfo{author}{S.~Aslam}, \bibinfo{author}{M.~P. Michaelides},
  \bibinfo{author}{H.~Herodotou},
\newblock \bibinfo{title}{Internet of ships: A survey on architectures,
  emerging applications, and challenges},
\newblock \bibinfo{journal}{IEEE Internet of Things journal}
  \bibinfo{volume}{7} (\bibinfo{year}{2020}) \bibinfo{pages}{9714--9727}.
\bibitem[{Yoo and Jeong(2022)}]{yoo2022vibration}
\bibinfo{author}{Y.~Yoo}, \bibinfo{author}{S.~Jeong},
\newblock \bibinfo{title}{Vibration analysis process based on spectrogram using
  gradient class activation map with selection process of cnn model and feature
  layer},
\newblock \bibinfo{journal}{Displays}  (\bibinfo{year}{2022})
  \bibinfo{pages}{102233}.
\bibitem[{Lei et~al.(2020)Lei, Yang, Jiang, Jia, Li, and
  Nandi}]{lei2020applications}
\bibinfo{author}{Y.~Lei}, \bibinfo{author}{B.~Yang},
  \bibinfo{author}{X.~Jiang}, \bibinfo{author}{F.~Jia},
  \bibinfo{author}{N.~Li}, \bibinfo{author}{A.~K. Nandi},
\newblock \bibinfo{title}{Applications of machine learning to machine fault
  diagnosis: A review and roadmap},
\newblock \bibinfo{journal}{Mechanical Systems and Signal Processing}
  \bibinfo{volume}{138} (\bibinfo{year}{2020}) \bibinfo{pages}{106587}.
\bibitem[{Li et~al.(2022)Li, Huang, Li, Liao, Chen, He, Yan, and
  Gryllias}]{li2022perspective}
\bibinfo{author}{W.~Li}, \bibinfo{author}{R.~Huang}, \bibinfo{author}{J.~Li},
  \bibinfo{author}{Y.~Liao}, \bibinfo{author}{Z.~Chen},
  \bibinfo{author}{G.~He}, \bibinfo{author}{R.~Yan},
  \bibinfo{author}{K.~Gryllias},
\newblock \bibinfo{title}{A perspective survey on deep transfer learning for
  fault diagnosis in industrial scenarios: Theories, applications and
  challenges},
\newblock \bibinfo{journal}{Mechanical Systems and Signal Processing}
  \bibinfo{volume}{167} (\bibinfo{year}{2022}) \bibinfo{pages}{108487}.
\bibitem[{Wang et~al.(2022)Wang, Wang, Ming, Zhang, Li, and Chu}]{wang2022semi}
\bibinfo{author}{X.~Wang}, \bibinfo{author}{T.~Wang},
  \bibinfo{author}{A.~Ming}, \bibinfo{author}{W.~Zhang},
  \bibinfo{author}{A.~Li}, \bibinfo{author}{F.~Chu},
\newblock \bibinfo{title}{Semi-supervised hierarchical attribute representation
  learning via multi-layer matrix factorization for machinery fault diagnosis},
\newblock \bibinfo{journal}{Mechanism and Machine Theory} \bibinfo{volume}{167}
  (\bibinfo{year}{2022}) \bibinfo{pages}{104445}.
\bibitem[{Chen et~al.(2022)Chen, Wu, Deng, Wang, and Wang}]{chen2022residual}
\bibinfo{author}{Z.~Chen}, \bibinfo{author}{J.~Wu}, \bibinfo{author}{C.~Deng},
  \bibinfo{author}{C.~Wang}, \bibinfo{author}{Y.~Wang},
\newblock \bibinfo{title}{Residual deep subdomain adaptation network: A new
  method for intelligent fault diagnosis of bearings across multiple domains},
\newblock \bibinfo{journal}{Mechanism and Machine Theory} \bibinfo{volume}{169}
  (\bibinfo{year}{2022}) \bibinfo{pages}{104635}.
\bibitem[{Wang et~al.(2021)Wang, Liu, Jiang, and Jiang}]{wang2021collaborative}
\bibinfo{author}{H.~Wang}, \bibinfo{author}{C.~Liu},
  \bibinfo{author}{D.~Jiang}, \bibinfo{author}{Z.~Jiang},
\newblock \bibinfo{title}{Collaborative deep learning framework for fault
  diagnosis in distributed complex systems},
\newblock \bibinfo{journal}{Mechanical Systems and Signal Processing}
  \bibinfo{volume}{156} (\bibinfo{year}{2021}) \bibinfo{pages}{107650}.
\bibitem[{Deng et~al.(2020)Deng, Diao, Wu, Zhang, Ma, and Zhong}]{deng2020high}
\bibinfo{author}{H.~Deng}, \bibinfo{author}{Y.~Diao}, \bibinfo{author}{W.~Wu},
  \bibinfo{author}{J.~Zhang}, \bibinfo{author}{M.~Ma},
  \bibinfo{author}{X.~Zhong},
\newblock \bibinfo{title}{A high-speed d-cart online fault diagnosis algorithm
  for rotor systems},
\newblock \bibinfo{journal}{Applied Intelligence} \bibinfo{volume}{50}
  (\bibinfo{year}{2020}) \bibinfo{pages}{29--41}.
\bibitem[{Zhang et~al.(2021)Zhang, Guan, Chen, Yang, Gong, and
  Yang}]{zhang2021adaptive}
\bibinfo{author}{Z.~Zhang}, \bibinfo{author}{C.~Guan},
  \bibinfo{author}{H.~Chen}, \bibinfo{author}{X.~Yang},
  \bibinfo{author}{W.~Gong}, \bibinfo{author}{A.~Yang},
\newblock \bibinfo{title}{Adaptive privacy-preserving federated learning for
  fault diagnosis in internet of ships},
\newblock \bibinfo{journal}{IEEE Internet of Things Journal}
  \bibinfo{volume}{9} (\bibinfo{year}{2021}) \bibinfo{pages}{6844--6854}.
\bibitem[{Iqbal et~al.(2019)Iqbal, Maniak, Doctor, and
  Karyotis}]{iqbal2019fault}
\bibinfo{author}{R.~Iqbal}, \bibinfo{author}{T.~Maniak},
  \bibinfo{author}{F.~Doctor}, \bibinfo{author}{C.~Karyotis},
\newblock \bibinfo{title}{Fault detection and isolation in industrial processes
  using deep learning approaches},
\newblock \bibinfo{journal}{IEEE Transactions on Industrial Informatics}
  \bibinfo{volume}{15} (\bibinfo{year}{2019}) \bibinfo{pages}{3077--3084}.
\bibitem[{Pan et~al.(2019)Pan, Chen, Zhou, Wang, and He}]{pan2019novel}
\bibinfo{author}{T.~Pan}, \bibinfo{author}{J.~Chen}, \bibinfo{author}{Z.~Zhou},
  \bibinfo{author}{C.~Wang}, \bibinfo{author}{S.~He},
\newblock \bibinfo{title}{A novel deep learning network via multiscale inner
  product with locally connected feature extraction for intelligent fault
  detection},
\newblock \bibinfo{journal}{IEEE Transactions on Industrial Informatics}
  \bibinfo{volume}{15} (\bibinfo{year}{2019}) \bibinfo{pages}{5119--5128}.
\bibitem[{Liu et~al.(2020)Liu, Guo, Al-Turjman, Muhammad, and
  de~Albuquerque}]{liu2020reliability}
\bibinfo{author}{S.~Liu}, \bibinfo{author}{C.~Guo},
  \bibinfo{author}{F.~Al-Turjman}, \bibinfo{author}{K.~Muhammad},
  \bibinfo{author}{V.~H.~C. de~Albuquerque},
\newblock \bibinfo{title}{Reliability of response region: a novel mechanism in
  visual tracking by edge computing for iiot environments},
\newblock \bibinfo{journal}{Mechanical systems and signal processing}
  \bibinfo{volume}{138} (\bibinfo{year}{2020}) \bibinfo{pages}{106537}.
\bibitem[{Kumar et~al.(2013)Kumar, Liu, Lu, and Bhargava}]{kumar2013survey}
\bibinfo{author}{K.~Kumar}, \bibinfo{author}{J.~Liu}, \bibinfo{author}{Y.-H.
  Lu}, \bibinfo{author}{B.~Bhargava},
\newblock \bibinfo{title}{A survey of computation offloading for mobile
  systems},
\newblock \bibinfo{journal}{Mobile networks and Applications}
  \bibinfo{volume}{18} (\bibinfo{year}{2013}) \bibinfo{pages}{129--140}.
\bibitem[{Mao et~al.(2017)Mao, You, Zhang, Huang, and Letaief}]{mao2017survey}
\bibinfo{author}{Y.~Mao}, \bibinfo{author}{C.~You}, \bibinfo{author}{J.~Zhang},
  \bibinfo{author}{K.~Huang}, \bibinfo{author}{K.~B. Letaief},
\newblock \bibinfo{title}{A survey on mobile edge computing: The communication
  perspective},
\newblock \bibinfo{journal}{IEEE communications surveys \& tutorials}
  \bibinfo{volume}{19} (\bibinfo{year}{2017}) \bibinfo{pages}{2322--2358}.
\bibitem[{Huda and Moh(2022)}]{huda2022survey}
\bibinfo{author}{S.~A. Huda}, \bibinfo{author}{S.~Moh},
\newblock \bibinfo{title}{Survey on computation offloading in uav-enabled
  mobile edge computing},
\newblock \bibinfo{journal}{Journal of Network and Computer Applications}
  (\bibinfo{year}{2022}) \bibinfo{pages}{103341}.
\bibitem[{Liao et~al.(2022)Liao, Lai, Yang, and Zeng}]{liao2022online}
\bibinfo{author}{L.~Liao}, \bibinfo{author}{Y.~Lai}, \bibinfo{author}{F.~Yang},
  \bibinfo{author}{W.~Zeng},
\newblock \bibinfo{title}{Online computation offloading with double
  reinforcement learning algorithm in mobile edge computing},
\newblock \bibinfo{journal}{Journal of Parallel and Distributed Computing}
  (\bibinfo{year}{2022}).
\bibitem[{Lu et~al.(2022)Lu, Mo, Feng, Gao, Zhao, Wu, and
  Nallanathan}]{lu2022secure}
\bibinfo{author}{W.~Lu}, \bibinfo{author}{Y.~Mo}, \bibinfo{author}{Y.~Feng},
  \bibinfo{author}{Y.~Gao}, \bibinfo{author}{N.~Zhao}, \bibinfo{author}{Y.~Wu},
  \bibinfo{author}{A.~Nallanathan},
\newblock \bibinfo{title}{Secure transmission for multi-uav-assisted mobile
  edge computing based on reinforcement learning},
\newblock \bibinfo{journal}{IEEE Transactions on Network Science and
  Engineering}  (\bibinfo{year}{2022}).
\bibitem[{Guo et~al.(2022)Guo, Zhao, Lai, Fan, Lei, and
  Karagiannidis}]{guo2022distributed}
\bibinfo{author}{Y.~Guo}, \bibinfo{author}{R.~Zhao}, \bibinfo{author}{S.~Lai},
  \bibinfo{author}{L.~Fan}, \bibinfo{author}{X.~Lei}, \bibinfo{author}{G.~K.
  Karagiannidis},
\newblock \bibinfo{title}{Distributed machine learning for multiuser mobile
  edge computing systems},
\newblock \bibinfo{journal}{IEEE Journal of Selected Topics in Signal
  Processing}  (\bibinfo{year}{2022}).
\bibitem[{Esposito et~al.(2017)Esposito, Castiglione, Pop, and
  Choo}]{esposito2017challenges}
\bibinfo{author}{C.~Esposito}, \bibinfo{author}{A.~Castiglione},
  \bibinfo{author}{F.~Pop}, \bibinfo{author}{K.-K.~R. Choo},
\newblock \bibinfo{title}{Challenges of connecting edge and cloud computing: A
  security and forensic perspective},
\newblock \bibinfo{journal}{IEEE Cloud computing} \bibinfo{volume}{4}
  (\bibinfo{year}{2017}) \bibinfo{pages}{13--17}.
\bibitem[{Liu et~al.(2020)Liu, Peng, Shou, Chen, and Chen}]{liu2020toward}
\bibinfo{author}{Y.~Liu}, \bibinfo{author}{M.~Peng}, \bibinfo{author}{G.~Shou},
  \bibinfo{author}{Y.~Chen}, \bibinfo{author}{S.~Chen},
\newblock \bibinfo{title}{Toward edge intelligence: Multiaccess edge computing
  for 5g and internet of things},
\newblock \bibinfo{journal}{IEEE Internet of Things Journal}
  \bibinfo{volume}{7} (\bibinfo{year}{2020}) \bibinfo{pages}{6722--6747}.
\bibitem[{Wu et~al.(2019)Wu, Huang, Xie, Nie, Bao, and Qin}]{wu2019ledge}
\bibinfo{author}{D.~Wu}, \bibinfo{author}{X.~Huang}, \bibinfo{author}{X.~Xie},
  \bibinfo{author}{X.~Nie}, \bibinfo{author}{L.~Bao}, \bibinfo{author}{Z.~Qin},
\newblock \bibinfo{title}{Ledge: Leveraging edge computing for resilient access
  management of mobile iot},
\newblock \bibinfo{journal}{IEEE Transactions on Mobile Computing}
  \bibinfo{volume}{20} (\bibinfo{year}{2019}) \bibinfo{pages}{1110--1125}.
\bibitem[{Cui et~al.(2020)Cui, Zhang, Zhang, Chen, Tao, and
  Zhang}]{cui2020online}
\bibinfo{author}{Q.~Cui}, \bibinfo{author}{J.~Zhang},
  \bibinfo{author}{X.~Zhang}, \bibinfo{author}{K.-C. Chen},
  \bibinfo{author}{X.~Tao}, \bibinfo{author}{P.~Zhang},
\newblock \bibinfo{title}{Online anticipatory proactive network association in
  mobile edge computing for iot},
\newblock \bibinfo{journal}{IEEE Transactions on Wireless Communications}
  \bibinfo{volume}{19} (\bibinfo{year}{2020}) \bibinfo{pages}{4519--4534}.
\bibitem[{Mao et~al.(2016)Mao, Zhang, and Letaief}]{mao2016dynamic}
\bibinfo{author}{Y.~Mao}, \bibinfo{author}{J.~Zhang}, \bibinfo{author}{K.~B.
  Letaief},
\newblock \bibinfo{title}{Dynamic computation offloading for mobile-edge
  computing with energy harvesting devices},
\newblock \bibinfo{journal}{IEEE Journal on Selected Areas in Communications}
  \bibinfo{volume}{34} (\bibinfo{year}{2016}) \bibinfo{pages}{3590--3605}.
\bibitem[{Barbarossa et~al.(2014)Barbarossa, Sardellitti, and
  Di~Lorenzo}]{barbarossa2014communicating}
\bibinfo{author}{S.~Barbarossa}, \bibinfo{author}{S.~Sardellitti},
  \bibinfo{author}{P.~Di~Lorenzo},
\newblock \bibinfo{title}{Communicating while computing: Distributed mobile
  cloud computing over 5g heterogeneous networks},
\newblock \bibinfo{journal}{IEEE Signal Processing Magazine}
  \bibinfo{volume}{31} (\bibinfo{year}{2014}) \bibinfo{pages}{45--55}.
\bibitem[{Zhang et~al.(2013)Zhang, Wen, Guan, Kilper, Luo, and
  Wu}]{zhang2013energy}
\bibinfo{author}{W.~Zhang}, \bibinfo{author}{Y.~Wen},
  \bibinfo{author}{K.~Guan}, \bibinfo{author}{D.~Kilper},
  \bibinfo{author}{H.~Luo}, \bibinfo{author}{D.~O. Wu},
\newblock \bibinfo{title}{Energy-optimal mobile cloud computing under
  stochastic wireless channel},
\newblock \bibinfo{journal}{IEEE Transactions on Wireless Communications}
  \bibinfo{volume}{12} (\bibinfo{year}{2013}) \bibinfo{pages}{4569--4581}.
\bibitem[{Zhang et~al.(2012)Zhang, Liu, Jiao, and Fu}]{zhang2012offload}
\bibinfo{author}{Y.~Zhang}, \bibinfo{author}{H.~Liu},
  \bibinfo{author}{L.~Jiao}, \bibinfo{author}{X.~Fu},
\newblock \bibinfo{title}{To offload or not to offload: An efficient code
  partition algorithm for mobile cloud computing},
\newblock in: \bibinfo{booktitle}{2012 IEEE 1st International Conference on
  Cloud Networking (CLOUDNET)}, \bibinfo{organization}{IEEE},
  \bibinfo{year}{2012}, pp. \bibinfo{pages}{80--86}.
\bibitem[{Mahmoodi et~al.(2016)Mahmoodi, Uma, and
  Subbalakshmi}]{mahmoodi2016optimal}
\bibinfo{author}{S.~E. Mahmoodi}, \bibinfo{author}{R.~Uma},
  \bibinfo{author}{K.~Subbalakshmi},
\newblock \bibinfo{title}{Optimal joint scheduling and cloud offloading for
  mobile applications},
\newblock \bibinfo{journal}{IEEE Transactions on Cloud Computing}
  \bibinfo{volume}{7} (\bibinfo{year}{2016}) \bibinfo{pages}{301--313}.
\bibitem[{Lu et~al.(2020)Lu, Gu, Luo, Ding, and Liu}]{lu2020optimization}
\bibinfo{author}{H.~Lu}, \bibinfo{author}{C.~Gu}, \bibinfo{author}{F.~Luo},
  \bibinfo{author}{W.~Ding}, \bibinfo{author}{X.~Liu},
\newblock \bibinfo{title}{Optimization of lightweight task offloading strategy
  for mobile edge computing based on deep reinforcement learning},
\newblock \bibinfo{journal}{Future Generation Computer Systems}
  \bibinfo{volume}{102} (\bibinfo{year}{2020}) \bibinfo{pages}{847--861}.
\bibitem[{Wang et~al.(2020)Wang, Tian, Cui, and Liu}]{wang2020reinforcement}
\bibinfo{author}{D.~Wang}, \bibinfo{author}{X.~Tian}, \bibinfo{author}{H.~Cui},
  \bibinfo{author}{Z.~Liu},
\newblock \bibinfo{title}{Reinforcement learning-based joint task offloading
  and migration schemes optimization in mobility-aware mec network},
\newblock \bibinfo{journal}{China Communications} \bibinfo{volume}{17}
  (\bibinfo{year}{2020}) \bibinfo{pages}{31--44}.
\bibitem[{Zhao et~al.(2020)Zhao, Wang, Xia, and Fan}]{zhao2020deep}
\bibinfo{author}{R.~Zhao}, \bibinfo{author}{X.~Wang}, \bibinfo{author}{J.~Xia},
  \bibinfo{author}{L.~Fan},
\newblock \bibinfo{title}{Deep reinforcement learning based mobile edge
  computing for intelligent internet of things},
\newblock \bibinfo{journal}{Physical Communication} \bibinfo{volume}{43}
  (\bibinfo{year}{2020}) \bibinfo{pages}{101184}.
\bibitem[{Ren et~al.(2020)Ren, Sun, and Peng}]{ren2020deep}
\bibinfo{author}{Y.~Ren}, \bibinfo{author}{Y.~Sun}, \bibinfo{author}{M.~Peng},
\newblock \bibinfo{title}{Deep reinforcement learning based computation
  offloading in fog enabled industrial internet of things},
\newblock \bibinfo{journal}{IEEE Transactions on Industrial Informatics}
  \bibinfo{volume}{17} (\bibinfo{year}{2020}) \bibinfo{pages}{4978--4987}.
\bibitem[{Min et~al.(2019)Min, Xiao, Chen, Cheng, Wu, and
  Zhuang}]{min2019learning}
\bibinfo{author}{M.~Min}, \bibinfo{author}{L.~Xiao}, \bibinfo{author}{Y.~Chen},
  \bibinfo{author}{P.~Cheng}, \bibinfo{author}{D.~Wu},
  \bibinfo{author}{W.~Zhuang},
\newblock \bibinfo{title}{Learning-based computation offloading for iot devices
  with energy harvesting},
\newblock \bibinfo{journal}{IEEE Transactions on Vehicular Technology}
  \bibinfo{volume}{68} (\bibinfo{year}{2019}) \bibinfo{pages}{1930--1941}.
\bibitem[{Hu et~al.(2018)}]{hu2018mobility}
\bibinfo{author}{R.~Q. Hu}, et~al.,
\newblock \bibinfo{title}{Mobility-aware edge caching and computing in vehicle
  networks: A deep reinforcement learning},
\newblock \bibinfo{journal}{IEEE Transactions on Vehicular Technology}
  \bibinfo{volume}{67} (\bibinfo{year}{2018}) \bibinfo{pages}{10190--10203}.
\bibitem[{Wei et~al.(2018)Wei, Zhao, Su, and Lu}]{wei2018dynamic}
\bibinfo{author}{Z.~Wei}, \bibinfo{author}{B.~Zhao}, \bibinfo{author}{J.~Su},
  \bibinfo{author}{X.~Lu},
\newblock \bibinfo{title}{Dynamic edge computation offloading for internet of
  things with energy harvesting: A learning method},
\newblock \bibinfo{journal}{IEEE Internet of Things Journal}
  \bibinfo{volume}{6} (\bibinfo{year}{2018}) \bibinfo{pages}{4436--4447}.
\bibitem[{Zhang et~al.(2020)Zhang, Du, Shen, and Wang}]{zhang2020dynamic}
\bibinfo{author}{J.~Zhang}, \bibinfo{author}{J.~Du}, \bibinfo{author}{Y.~Shen},
  \bibinfo{author}{J.~Wang},
\newblock \bibinfo{title}{Dynamic computation offloading with energy harvesting
  devices: A hybrid-decision-based deep reinforcement learning approach},
\newblock \bibinfo{journal}{IEEE Internet of Things Journal}
  \bibinfo{volume}{7} (\bibinfo{year}{2020}) \bibinfo{pages}{9303--9317}.

\end{thebibliography}

\clearpage

\end{document}